\documentclass[useAMS,usenatbib]{mn2e}
\usepackage{amsmath}
\usepackage{amsfonts}
\usepackage{graphics}
\usepackage{subfigure}
\usepackage{epsfig}
\usepackage{rotate}
\usepackage{ifpdf}
\usepackage{graphicx,times}
\usepackage{natbib}
\usepackage{amssymb}
\usepackage{lipsum}

\title[Chaos and dynamical trends in barred galaxies]
{Chaos and dynamical trends in barred galaxies: bridging the gap between
$N$-body simulations and time-dependent analytical models}

\author[Manos $\&$ Machado]{T.~Manos$^{\mathrm{1},\mathrm{2}}
\thanks{E-mail: thanos.manos@uni-mb.si}$ and Rubens E.~G. Machado$^{\mathrm{3}}$\\
$^1$ CAMTP - Center for Applied Mathematics and Theoretical Physics, University of Maribor, Krekova 2, SI-2000 Maribor, Slovenia.\\
$^2$ School of Applied Sciences, University of Nova Gorica, Vipavska 11c, SI-5270 Ajdov\v s\v cina, Slovenia.\\
$^3$ Instituto de Astronomia, Geof\'isica e Ci\^encias Atmosf\'ericas, Universidade de S\~ao Paulo, R. do Mat\~ao 1226, 05508-090 S\~ao Paulo, Brazil.}

\date{Received 2013 November 17}
\pubyear{2013}

\def\LaTeX{L\kern-.36em\raise.3ex\hbox{a}\kern-.15em
    T\kern-.1667em\lower.7ex\hbox{E}\kern-.125emX}

\begin{document}
\maketitle

\begin{abstract}
Self-consistent $N$-body simulations are efficient tools to study galactic
dynamics. However, using them to study individual trajectories (or ensembles)
in detail can be challenging. Such orbital studies are important to shed light
on global phase space properties, which are the underlying cause of observed
structures. The potentials needed to describe self-consistent models are
time-dependent. Here, we aim to investigate dynamical properties
(regular/chaotic motion) of a non-autonomous galactic system, whose
time-dependent potential adequately mimics certain realistic trends arising
from $N$-body barred galaxy simulations. We construct a fully time-dependent
analytical potential, modeling the gravitational potentials of disc, bar and
dark matter halo, whose time-dependent parameters are derived from a
simulation. We study the dynamical stability of its reduced time-independent
2-degrees of freedom model, charting the different islands of stability
associated with certain orbital morphologies and detecting the chaotic and
regular regions. In the full 3-degrees of freedom time-dependent case, we show
representative trajectories experiencing typical dynamical behaviours, i.e.,
interplay between regular and chaotic motion for different epochs. Finally, we
study its underlying global dynamical transitions, estimating fractions of
(un)stable motion of an ensemble of initial conditions taken from the
simulation. For such an ensemble, the fraction of regular motion increases with
time.
\end{abstract}


\section{Introduction} \label{intro}

Orbits are the fundamental building blocks of any galactic structure and their
properties give important insight for understanding the formation and evolution
of such structures \citep{BinneyTremaine}. Our understanding relies
significantly on the adequacy and efficiency of the models used either in the
time-dependent (TD) self-consistent models or in the rather `simpler'
analytical time-independent (TI) ones. The presence of chaos, manifested as
unstable orbital motion, and expressed by exponential divergence of nearby
trajectories, is broadly studied over the past years \citep[for a good review
on chaos in galaxies, see the book by][]{ContBook2002}. It is by now well
accepted that the chaotic or regular nature of orbits influences the general
stability of the $N$-body simulations, which is straightforwardly related to
the underlying dynamics. Therefore, studying the general stability and the
detailed structure of the phase (but also of the configuration) space of
analytical models can be proven to be very useful, provided that these
potentials are realistic in terms of representing density distribution profiles
close to those derived by simulations.

The general nature of an orbit, in conservative (TI) systems, can only be one
of the following: periodic (stable or unstable), quasi-periodic or chaotic
\citep{LichLieb}. Nevertheless, there are cases where chaos can be
characterized as \textit{weak}, suggesting that orbits spend a significant
fraction of their time in confined regimes and do not fill up phase space as
`homogeneously' as the \textit{strongly} chaotic ones. In these case, the
different rate of diffusion in the phase space plays an important role,
associated for example the weak chaotic motion  with barred or spiral galaxy
features, giving rise to a number of interesting results \citep[see
e.g.][]{AthaRomMas2009MNRAS,AthaRomBosMas2010MNRAS,AthRomBosMas2009MNRAS,
HarKalMNRAS2009,HarKalCont2011MNRAS,HarKalCont2011IJBC,ConHar2013MNRAS,
KauCont1996A&A,PatAthQui1997ApJ,Pat2006MNRAS,RomMasAthGar2006A&A,RomAthMasGar2007A&A,
TsoKalEftCon2009A&A,BruChiPfe2011A&A,BouManAntCeMDA2012}. There are also
several results in the recent literature showing that strong local instability
does not necessarily imply widespread diffusion in phase space
\citep{CacCinFer2010CeMDA,GioCin2004A&A}. In
\cite{ConHar2008IJBC,ConHar2010CeMDA} `stickiness' was studied thoroughly in
2-degrees of freedom (d.o.f.) while in \cite{KatPatPin2011IJBC},
\cite{KatPat2011IJBC}, \cite{KatPatCon2011IJBC} and in \cite{ManSkoAnt2012IJBC}
the role of `sticky' chaotic orbits and the diffusive behavior, in the
neighborhood of invariant tori surrounding periodic solutions of the
Hamiltonian in the vicinity of periodic orbits in conservative systems, was
also studied.

Doubtless, the different rate of diffusion consists in a very important topic
when studying chaotic motion in galactic systems. However, in this paper we
mainly focus and concentrate on more general dynamical (in)stability trends
using rather standard chaos detection techniques. Over the last years, many
chaos detection methods have been developed and also compared, exploiting, in
general, either the tangent dynamics of an orbit under study via the
simultaneous evolution of its deviation vectors or the analysis of time series
constructed by its coordinates. A complete list of all these techniques can be
found, e.g. in \cite{ContBook2002}, in a review paper \citep{SkoLNP2010} and
more recently in \cite{MafDarCinGio2011CeMDA,MafDarCinGio2013MNRAS} and
references therein. Their efficiency and accuracy on the distinction between
regular and chaotic motion has been thoroughly studied and discussed therein as
well.

Recently, in \cite{ManBouSkoJPhA2013}, a study was carried out focusing on the
dynamics of a barred galaxy model containing a disc and a bulge component.
Considering a TD analytical model -- extending a TI one \citep{ManAthMNRAS2011}
-- whose mass parameters of the bar and disc potential vary linearly as
functions of time (the one at expense of the other). Two very general
conceivable cases in barred galaxies were analyzed: (a) a model where the mass
of bar grows, considering a common trend found in $N$-body simulations due to
the exchange of angular momentum \citep[see
e.g.][]{AthMisMNRAS2002,AthMNRAS2003} and (b) a case where the bar gets weaker
by losing mass \citep[see e.g.][]{Combes:2008,CombMSAIS2011}. There, a new
reliable way of using the Generalized Alignment Index (GALI) chaos detection
method was used for estimating the relative fraction of chaotic
\textit{vs.}~regular orbits in such TD potentials. We stress here that in the
TD models, individual trajectories may display sudden transitions from regular
to chaotic behavior and vice versa during their time evolution and in general
the `sticky' behaviour, as discussed in the literature, is less pronounced.
This is also the typical case in the $N$-body simulations where, generally
speaking, the motion may also be either: (i) regular throughout the whole
evolution, (ii) chaotic throughout the whole evolution, (iii) alternate between
chaotic and regular motion with simultaneously orbital shape change (but not
necessarily), e.g. from disc to bar like, etc...

Completing furthermore the picture on previous studies on the study of
(ir)regular motion in TD systems proposed for cosmological and galactic models,
we could refer to a number of publications starting with the definition of the
orbital complexity $n(k)$ of an orbital segment which corresponds to the number
of frequencies in its discrete Fourier spectrum that contain a $k$-fraction of
its total power, used by \cite{KandEckBra1997A&A,SioEckKand1998NYASA}. The
quantity $n(k)$ was later associated with the short-time evolution of the
Lyapunov Exponents (LEs) for TD models in \citep{SioEckKand1998NYASA}. A study
in a cosmological system where trajectories were changing their dynamical
nature (from regular to chaotic and/or vice-versa) was performed by
\citet{KandDru1998NYASA}. The role of friction, noise, periodic driving, black
holes  was studied by \citet{SioKand2000MNRAS} while later on this was extended
\citep[in][]{KandVasSid2003MNRAS,TerKand2004MNRAS} following the transient
chaos due to damped oscillations. In \cite{Sid2009chas.book} an exponential
function of time was added to the H\'{e}non-Heiles potential, using the
so-called `pattern method' as chaos detection tool and more recently  the
dynamics of some simple TD galactic models was investigated
\citep[in][]{CarPap2003A&A,Zot2012NewA}.

\begin{figure*}
\begin{center}
\includegraphics[width=\textwidth]{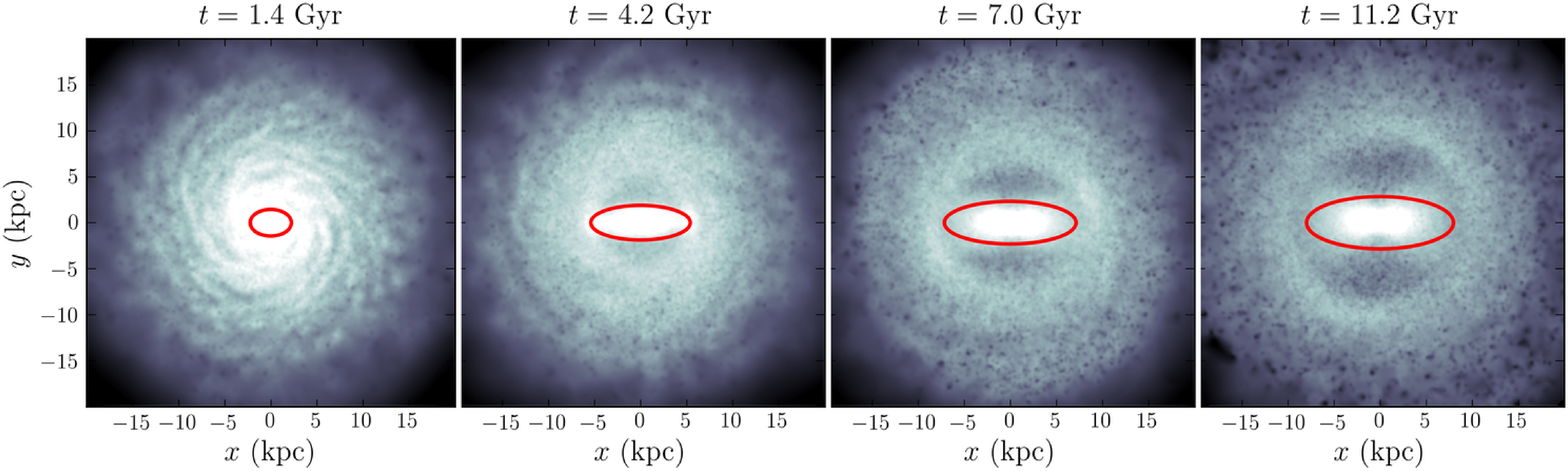}\\
\includegraphics[width=\textwidth]{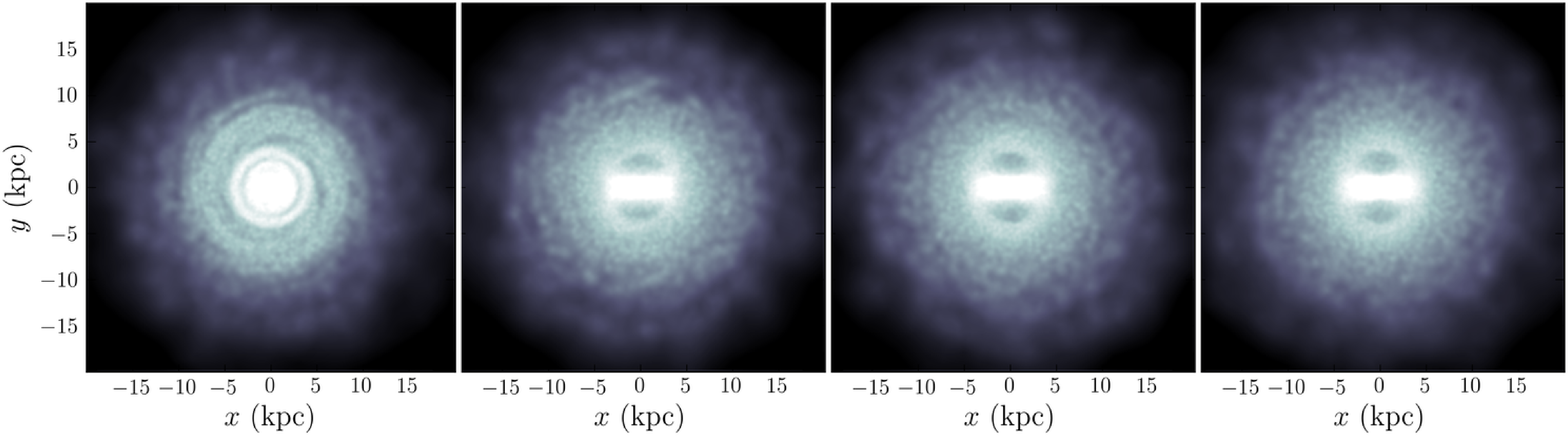}
\caption{(Colour online) Snapshots of the $N$-body simulation (upper panels) at
four different times, displaying stellar density, on the same range, projected
on the $xy$ plane. Each frame is 40 by 40 kpc. To illustrate bar lengths and
shapes, we overlay ellipses (which are not isophotal fits). Rather, their
semi-major axes are obtained from the radial $m=2$ Fourier component of the
mass distribution, and from shape measurements via the inertia tensor (see
text). The lower panels display the result of evolving an ensemble of initial
conditions in the presence of the constructed TD analytical potential.}
\label{fig:frames}
\end{center}
\end{figure*}

Regarding the galaxies' evolution and formation of their several features, it
is generally accepted that the most appropriate way to study them is by
analyzing $N$-body simulations. The self-consistency of the models in this
approach captures much better several details of the general dynamics. The
direct application of chaos detection methods to individual orbits is still a
rather difficult task while, for a large ensemble of particles, it is even
harder if not unfeasible. To overcome this obstacle, mean field potentials have
been used in the literature in order to study in more detail the dynamical
properties of a specific $N$-body simulation. These potentials are referred to
as `frozen' and they are TI and are derived at specific snapshots of the
simulations. Hence, one can apply chaos detection tools to the mean field
potential instead of the $N$-body simulation. For example,
\citet{MuzCarWax2005CeMDA} used an elliptical galaxy simulation (no bar or
halo) without dissipation which collapses and eventually reaches an equilibrium
state. Then, by taking a quadrupolar expansion of the frozen snapshot, they
derive a stationary smooth potential. In \cite{VogStaKal2006MNRAS} and
references therein, the authors deal with disc galaxies, focusing mainly on the
spiral structures rather than bars (no halo) while the extraction of the mean
field potential is again performed in a similar manner. Following this
approach, the role of chaotic motion and diffusion rate in barred spiral
galaxies has also been studied \citep{HarKalMNRAS2009,HarKalCont2011MNRAS,
HarKalCont2011IJBC,MafDarCinGio2013MNRAS,ConHar2013MNRAS} while some
applications to the Milky Way bar can be found in
\cite{WangZhaoMaoRich2012MNRAS} and recently a new code for orbit analysis and
Schwarzschild modelling of triaxial stellar systems was given in
\cite{Vas2013MNRAS}. Nevertheless, following this approach one only derives a
stationary mean field model for an equilibrium state of a simulation under
study. Furthermore, it does not incorporate an appropriate type set of
parameters that would be able to describe and reproduce the time-dependencies
in axis ratios, masses, pattern speed, etc.~of the several components of a
model, like for example the growth of the bar component or the evolution of the
disc in time. Let us point out here, the fact that in all these approaches the
orbits under study can be only either regular or chaotic. The latter ones may
be further distinguished to strongly or weakly chaotic, depending on their
diffusion properties, sticky effects etc... during the whole evolution.

In this paper, we consider an $N$-body simulation of a disc galaxy embedded in
a live halo (i.e. both the stellar disc and the dark matter halo are
represented by responsive particles). Disc-halo interaction leads to the
formation of a strong bar. We then measure how the galaxy components vary in
time during the simulation. The time evolution of the structural parameters is
provided as input to the analytical TD model we build. This `candidate'
analytical mean field potential is meant to mimic the $N$-body simulation
evolution and more importantly to generate orbits with more similar (and in
some sense `richer') morphological behaviour to those of the $N$-body
simulation, i.e., permitting for individual orbits the interplay between
regular and chaotic epochs as time evolves and providing a stable structure at
the same time. Note that, in TI frozen models an orbit cannot convert from
chaotic to regular. Our TD model is composed of three components (bar, disc and
halo) whose parameters were fitted with the $N$-body measurements, chiefly via
the rotation curves. Note that many simplifying assumptions are made. For
example, our TD model considers an (ellipsoidal) analytical bar component which
is not always an excellent approximation of the shape of the actual $N$-body
bar. Likewise, the analytical description of the halo and the disc cannot be
expected to behave identically, either. However, our goal is to study the
general dynamical impact in stability caused by the bar's growth in time (as it
happens in the $N$-body simulation). Thus, by using a realistic TD model,
without aiming to describe of the exact detailed dynamics yielding from the
simulation, we can use chaos detection tools and quantify general trends of
relative regular and chaos in the phase and configuration space. Keeping this
in mind, we draw (disc) initial conditions directly from the simulation and we
evolve in time them with the mean field TD potential.

The paper is organized as follows: In Section~\ref{NbodyvsPot} we describe the
$N$-body simulation we used, and we already start presenting the first part of
our results, on the construction of a novel time-dependent analytical model and
we present its dynamical properties with respect to this simulation. In
Section~\ref{tools} we give briefly the definitions of the chaos detection
methods employed and their behaviors for chaotic and regular motion. In
Section~\ref{2DTImodel} we explore the global phase space dynamics of the
derived analytical model, first for the stationary case (frozen potential),
e.g. for different (but fixed) sets of parameters which correspond in a sense
to different snapshots of the simulation. In Section~\ref{TDmodel} we study the
dynamical trends and stability of the 3-d.o.f. time-dependent analytical model
for both a sample of single orbits but also for a large ensemble of initial
conditions from the simulation. Finally, in Section~\ref{concl} we summarize
our findings.

\section{$N$-body simulation and construction of the analytical model}
\label{NbodyvsPot}

Our aim is to construct a time-dependent analytical model that represents the
evolution of a barred galaxy. In order to have an astrophysically well
motivated model, we supply it with (time-dependent) structural parameters
measured from the output of an $N$-body simulation.

\subsection{The $N$-body simulation}

To serve as the base reference for the analytical model, we use one of the
simulations described in \cite{MachadoAthanassoula2010}. For simplicity, we
select initial conditions with a spherical halo. The mass of the  stellar disc
is $M_{d}=5\times 10^{10}~M_{\odot}$, with an exponential density profile of
radial scale length $R_{d}=3.5$~kpc, and vertical scale height $z_{0}=0.7$~kpc:
\begin{equation} \label{rho_disc}
\rho_{d}(R,z) = \frac{M_{d}}{4 \pi z_{0} R_{d}^{2}}
\exp{\left(-\frac{R}{R_{d}}\right)}
\mathrm{sech}^{2}{\left(\frac{z}{z_{0}}\right)},
\end{equation}
The spherical dark matter halo has a \cite{Hernquist1993} density profile and it is five times
more massive than the disc:
\begin{equation}
\rho_{h}(r) = \frac{M_{h}}{2 \pi^{3/2}} \frac{\alpha}{r_{c}}
\frac{\exp{(-r^{2}/r_{c}^{2})}}{r^{2}+\gamma'^{2}},
\end{equation}
where $M_h=2.5\times 10^{11}~M_{\odot}$ is the mass of the halo, $\gamma'=1.7$~kpc is a core radius and $r_{c}=35$~kpc is a cutoff radius. The normalisation constant $\alpha$ is defined by
\begin{equation}
\alpha = \{ 1 - \sqrt{\pi} q \exp{(q^{2})} [1- \textrm{erf}(q)] \}^{-1}
\end{equation}
where $q=\gamma' / r_{c}$. For additional details on the initial conditions,
see \cite{MachadoAthanassoula2010}.

This is a fairly representative collisionless simulation of a strongly barred
galaxy. Four snapshots of the disc particles are displayed in the upper row of
Fig.~\ref{fig:frames}. It was performed with the $N$-body code
\textsc{gyrfalcon} \citep{Dehnen2000, Dehnen2002} using a total of 1.2 million
equal-mass particles, with a gravitational softening length of 0.175~kpc,
resulting in 0.1 per cent energy conservation. The simulation was carried out
for approximately one Hubble time.

$N$-body simulations have been employed to study chaotic motion in simplified
models of disc galaxies. For example, \cite{VogStaKal2006MNRAS} study chaos and
spiral structure in rotating disc galaxies, but those galaxies are not embedded
in dark matter haloes.

The connection between chaos and bars was also analysed by
\cite{ElZantShlosman2002}, with models were set up by the addition of disc, bar
and halo components. They found that in centrally concentrated models, even a
mildly triaxial halo lead to the onset of chaos and the dissolution of the bar
in a timescale shorter than the Hubble time.

\subsection{The time-dependent analytical model}

We construct an analytical model that is described by its total gravitational
potential \mbox{$V = V_{B}(t) + V_{D}(t) + V_{H}(t)$}, where the three
components correspond to the potentials of the bar, disc, and halo,
respectively. These components will evolve in time, in accordance with the
behaviour we measure from the simulation. Each of these three components is
represented in the following way:
\begin{figure}
\begin{center}
\includegraphics[scale=0.6]{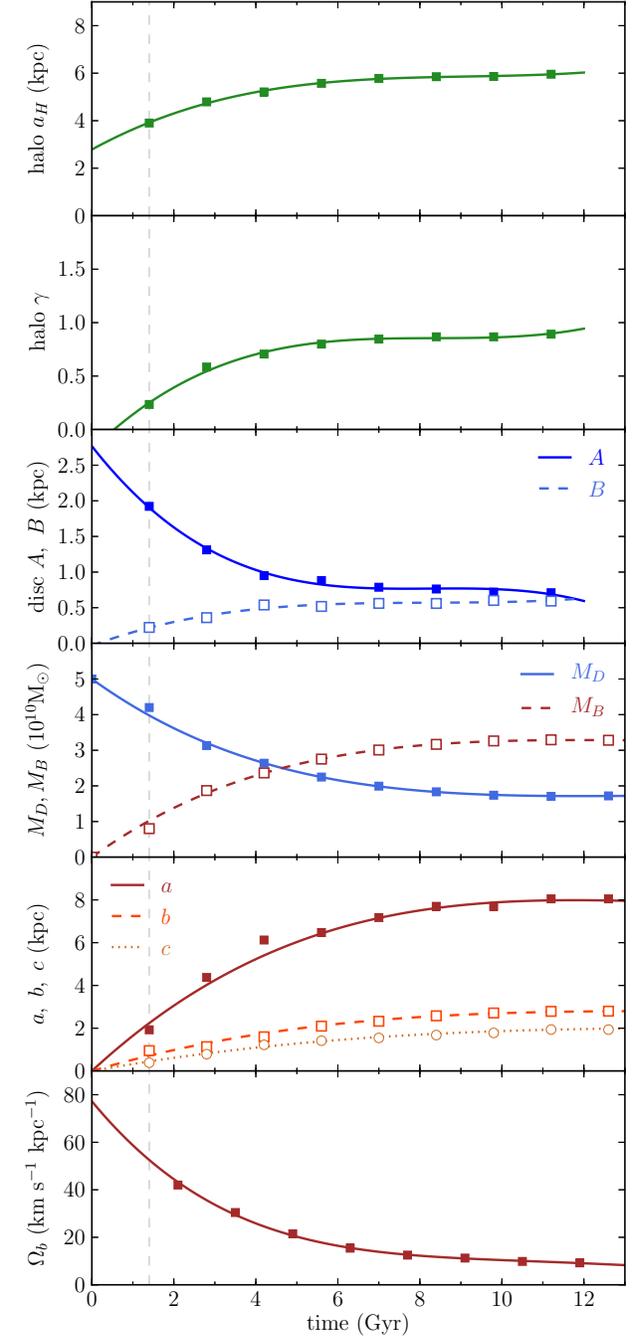}
\caption{(Colour online) Time evolution of the halo, disc and bar parameters,
measured from the $N$-body simulation (points) and supplied to the analytical
model (fitted polynomials). First and second panel: parameters of the Dehnen
halo profile. Third: parameters of the Miyamoto--Nagai disc. Fourth: bar mass
and disc mass. Fifth: semi-major axes of the bar. Sixth: bar pattern speed.}
\label{fig:poly}
\end{center}
\end{figure}

\begin{enumerate}
\item[\bf(i)] A triaxial Ferrers bar \citep{Fer1877}, whose density is given by:
\begin{eqnarray}
  \rho(x,y,z) = \left\{
  \begin{array}{l l}
    \rho_{c}(1-m^{2})^{2}& \quad \textrm{if} \quad m<1,\\
    \quad 0& \quad \textrm{if} \quad m \geq 1,\\
  \end{array} \right.
\end{eqnarray}
where $\rho_{c}=\frac{105}{32\pi}\frac{G M_{B}(t)}{abc}$ is the central
density, $M_{B}(t)$ is the mass of the bar, which changes in time, and
$m^{2}=\frac{x^{2}}{a^{2}}+\frac{y^{2}}{b^{2}}+\frac{z^{2}}{c^{2}}$,
\mbox{$a>b>c> 0$}, with $a,b$ and $c$ being the semi-axes of the ellipsoidal
bar. The corresponding bar potential is:
\begin{equation}\label{Ferr_pot}
    V_{B}(t)= -\pi Gabc \frac{\rho_{c}}{3}\int_{\lambda}^{\infty}
    \frac{du}{\Delta (u)} (1-m^{2}(u))^{3},
\end{equation}
where $G$ is the gravitational constant (set to unity),
$m^{2}(u)=\frac{x^{2}}{a^{2}+u}+\frac{y^{2}}{b^{2}+u}+\frac{z^{2}}{c^{2}+u}$,
$\Delta^{2} (u)=({a^{2}+u})({b^{2}+u})({c^{2}+u})$, and $\lambda$ is the unique
positive solution of $m^{2}(\lambda)=1$, outside of the bar ($m \geq 1$), while
$\lambda=0$ inside the bar. The analytical expression of the corresponding
forces are given in \cite{PfeA&A1984a}. In our model, the shape parameters
(i.e. the lengths of the ellipsoid axes $a$, $b$ and $c$ are) are also functions of
time.

\item[\bf (ii)] A disc, represented by the Miyamoto--Nagai potential
  \citep{MNPASJ1975}:
 \begin{equation}\label{eq:MNPot}
   V_{D}(t)=- \frac{GM_{D}(t)}{\sqrt{x^{2}+y^{2}+(A+\sqrt{z^{2}+B^{2}})^{2}}},
\end{equation}
where $A$ and $B$ are its horizontal and vertical scale-lengths, and $M_{D}(t)$
is the mass of the disc. Here, `disc mass', refers to the stellar mass
excluding the bar. As the bar grows, its mass increases at the expense of the
remainder of the disc mass, such that the total stellar mass is constant:
$M_{B}(t)+M_{D}(t) = 5\times 10^{10}~M_{\odot}$. The parameters $A$ and $B$ are
also functions of time.

\item[\bf (iii)] A spherical dark matter halo, represented by a Dehnen potential \citep{DehnenMNRAS1993}:
\begin{equation}\label{eq:DehnenPot}
V_{H}(t)=\frac{GM_{H}}{a_H} \times\left\{
                      \begin{array}{lll}
                        -\frac{1}{2-\gamma}\left[1-\left(\frac{r}{r+a_H}\right)^{2-\gamma}\right]&, & \hbox{$\gamma \neq 2$,} \\
                        \ln \frac{r}{r+a_H}&, & \hbox{$\gamma = 2$.}
                      \end{array}
                    \right.
\end{equation}
$M_{H}$ is the halo mass, $a_{H}$ is a scale radius and the dimensionless
parameter $\gamma$ (within $ 0 \leq \gamma < 3$) governs the inner slope. The
halo mass is constant throughout, but the parameters $a_{H}$ and $\gamma$ are
functions of time. For $\gamma <2$ its finite central value is equal to
$(2-\gamma)^{-1}GM_H/a_H$.
\end{enumerate}

\begin{figure*}
\begin{center}
\includegraphics[width=\textwidth]{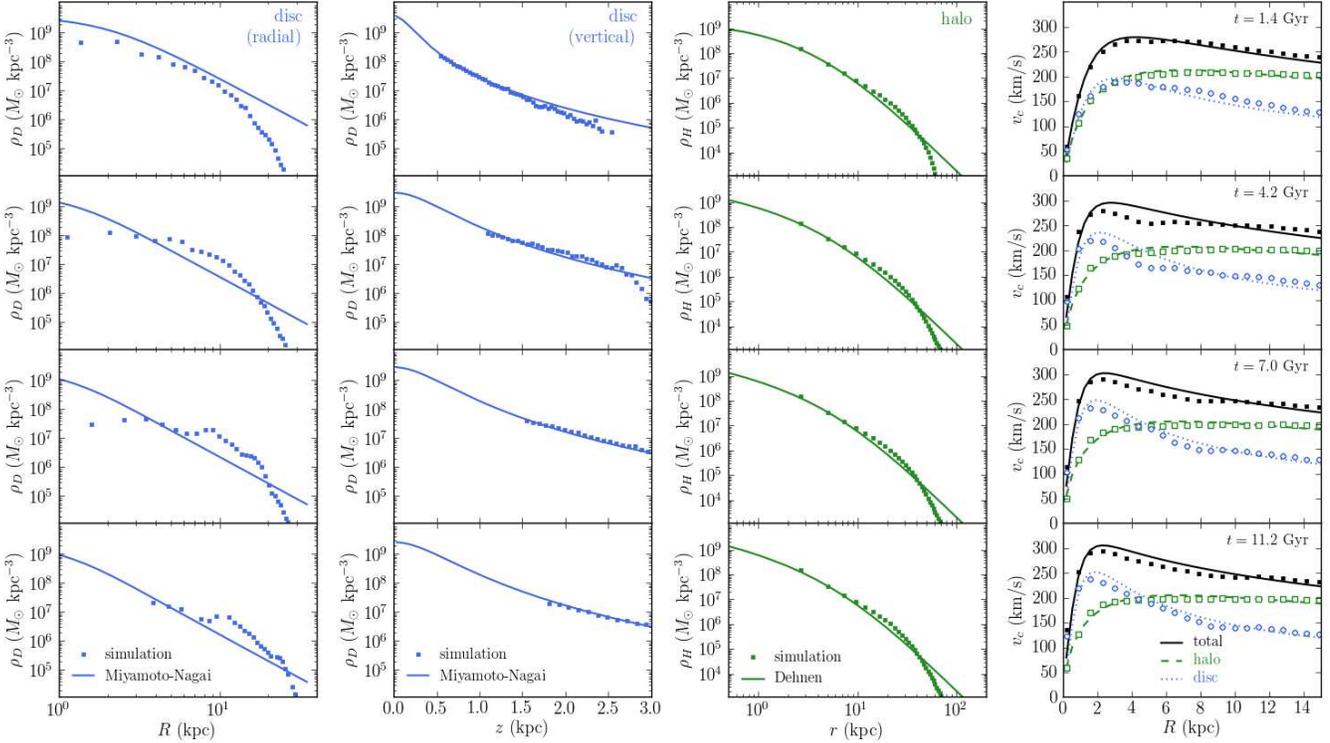}
\caption{(Colour online) Parameters of the Dehnen halo profile and of the
Miyamoto--Nagai disc are obtained by fitting the circular velocity curves at
each time. Here we display four different times. The fourth column exhibits
total, halo and disc circular velocity curves. The first and second columns
show the disc density profiles (radial and vertical, respectively). The third
column has the halo density profile. Points come from measurements of the
$N$-body simulation, while lines are fitted profiles.} \label{fig:vcfits}
\end{center}
\end{figure*}

Instead of attempting to use the (disc and halo) profiles from the $N$-body
simulations, we opted to represent the bar, disc and halo using respectively
the Ferrers, Miyamoto--Nagai and Dehnen profiles. There are two reasons for
such a choice. First, our approach requires analytical simplicity that could
not be afforded by the profiles used in the initial conditions of the numerical
simulation. Secondly, due to bar formation and evolution, the initial disc
profile in the simulation soon becomes a poor representation in the inner part
of the galaxy, where the bar resides. In this sense, it is not advantageous to
continue using the initial profiles to model later times. A Miyamoto--Nagai
disc provides a sufficient approximation for our purposes. Likewise, even
though the \cite{Hernquist1993} halo profile is well suited for numerical
purposes, it is inconvenient from the analytical point of view. We experimented
with simple logarithmic halo profiles (because their rotation curves are also
appropriate), but the \cite{DehnenMNRAS1993} profile was preferable, as it
provided equally acceptable rotation curves, and a more satisfactory global
approximation of the mass distribution. Similarly, fitting the bar by a Ferrers
ellipsoid is a justifiable approximation. Surely, it fails to capture the
$N$-body bar in all its complexity, particularly after the buckling
instability, when the bar is substantially strong and develops the
peanut-shaped feature. In general, the $N$-body bar will be more boxy than the
Ferrers shape would allow. Nevertheless, fitting an ellipsoid of the same
extent allows us to obtain plausible shapes, to determine the bar orientation
and to estimate its mass adequately. Ultimately, regardless of small deviations
in the density profiles, our goal is to obtain an analytical total potential
that is approximately comparable to the overall potential of the simulation.

From the simulation, we are able to measure several quantities as a function
of time, which are then used to inform the analytical model.

For the bar, the required parameters are the bar mass, the bar shape and the
bar pattern speed $\Omega_{b}$. First, we estimate the bar length as a function
of time. This is done by measuring the relative contribution of the $m=2$
Fourier component of the mass distribution as a function of radius, for each
time step, and finding the radius at which the $m=2$ has its most intense drop
after the peak. This radius $a$ is associated with the bar length. Then we
estimate the bar shape by calculating the axis ratios $b/a$ and $c/a$ from the
eigenvalues of the inertia tensor in this region. The bar mass is measured by
simply adding up the mass enclosed within an ellipsoid of axes $a,b,c$.
Finally, the successive orientations of the major axis as a function of time
are used to compute the pattern speed. The resulting time evolution of all
these quantities are displayed in the fourth, fifth and sixth panels of
Fig.~\ref{fig:poly}. Each of these parameters is measured at several time
steps, a sample of which is shown, along with the resulting polynomial fits.

The disc mass $M_{D}(t)$ is known once the bar mass has been measured, and the
halo mass $M_{H}$ is constant. One still requires the time evolution of two
disc parameters ($A$, $B$) and two halo parameters ($a_{H}$, $\gamma$). This is
achieved by measuring the rotation curves directly from the simulation (at each
time step), and then fitting the analytical $v_{c}(R)$ to these data. Since the
disc and halo potentials are known from equations (\ref{eq:MNPot}) and
(\ref{eq:DehnenPot}), we obtain their respective analytical circular velocities
from $v_{c}^2 = R \frac{dV}{dR}$:

\begin{eqnarray}
v_{c,D}^{2}(R) &=& R^2 ~ \frac{G M_D}{\left[ R^2 + (A+B)^2\right]^{3/2}} \label{eq:vcD}  \\
v_{c,H}^{2}(R) &=& G M_H ~ \frac{r^{2-\gamma}}{(r+a_{H})^{3-\gamma}}\label{eq:vcH}
\end{eqnarray}

Fitting Eq.~(\ref{eq:vcH}) to the measured halo rotation curve, we obtain
$a_{H}$ and $\gamma$. In the case of the disc, it is not enough to fit
Eq.~(\ref{eq:vcD}). One must simultaneously fit the Miyamoto--Nagai density
profile to disambiguate the $A+B$ (ignoring the inner part of the disc). When
fitting the disc rotation curve, we assume the total stellar mass (i.e. we take
both disc and bar mass into account). Since the circular velocities rely on
azimuthally averaged quantities, the presence of the bar does not greatly
interfere with the quality of the fits, while its removal would lead to
spurious results. The measured rotation curves (disc, halo and total), as well
as the resulting fitted circular velocities, are displayed in the fourth column
of Fig.~\ref{fig:vcfits} (at four illustrative instants in time).  Errors in
the fitted parameters of rotation curves were typically of about 5 per cent or
less. Also shown in the first, second and third panels of
Fig.~\ref{fig:vcfits}, are the disc (radial and vertical) and halo density
profiles. The points correspond to simulation measurements and the lines give
the resulting fits.

One of the main arguments in favor of the adequacy of our analytical model is
evidenced by the fact that its total rotation curves are in good agreement with
those measured from the simulation. This indicates that the choices of profiles
were not unreasonable, as they result in a globally similar gravitational
potential. Even if individually the densities of the components are idealized
simplifications, the similarity of the total potential ensures that the overall
dynamical evolution should be sufficiently well approximated.

Finally, the resulting time evolution of the halo and disc structural
parameters, measured in the manner described above, are displayed in the first
to fourth panels of Fig.~\ref{fig:poly}. With these, the time-dependence of the
analytical model is fully specified. In Table~\ref{tab1} we summarize the
analytical model by showing a sample of parameters for the Ferrers bar,
Miyamoto--Nagai disc and Dehnen halo potential as fitted by the $N$-body
simulation, at four times.

\begin{table*}
\centering \footnotesize \caption{The parameters for the Ferrers bar,
Miyamoto--Nagai disc and Dehnen halo potential as fitted by the $N$-body
simulation, at four times.}
\label{tab1}
\begin{tabular}{c c c c c c c c c c c c c c c}
\hline
 & & & & Bar & & & & & Disc & & & & Halo & \\
\hline
  time  & & $a$ & $b$ & $c$ & $\Omega_b$ & $M_B$ & &$A$ & $B$ & $M_D$ & & $a_H$ & $\gamma$ & $M_H$ \\
 (Gyr)  & & (kpc) & (kpc) & (kpc) & (km s$^{-1}$ kpc$^{-1}$) & ($10^{10} M_{\odot}$) & & (kpc) & (kpc) & ($10^{10} M_{\odot}$) & & (kpc) & &  ($10^{10} M_{\odot}$)  \\
\hline
~1.4 & & 2.24 & 0.71 & 0.44 & 52 & 1.40 & & 1.92 & 0.22 & 3.96 & & 3.90 & 0.23 & 25 \\
~4.2 & & 5.40 & 1.76 & 1.13 & 24 & 2.36 & & 0.95 & 0.53 & 2.64 & & 5.21 & 0.71 & 25 \\
~7.0 & & 7.15 & 2.38 & 1.58 & 14 & 3.02 & & 0.78 & 0.56 & 1.98 & & 5.77 & 0.85 & 25 \\
11.2 & & 7.98 & 2.76 & 1.93 & ~9 & 3.30 & & 0.71 & 0.59 & 1.70 & & 5.95 & 0.89 & 25 \\
  \hline
\end{tabular}
\end{table*}

\subsection{Bar strength}

In order to measure the bar strengths in analytical models,
\cite{ManAthMNRAS2011} had employed the $Q_b$ parameter
\citep{ButBloKna2003AJ,ButLauSal2004AJ}, which is a measure of the relative
strength of the non-axisymmetric forces. Here, we opt instead to use a method
more familiar to $N$-body simulations, namely measurements of the $m=2$ Fourier
component of the mass distribution. For the $N$-body simulation, we measure
this component straightforwardly as a function of radius and then take the
maximum amplitude to be the $A_2$ \citep*[see e.g.][]{Athanassoula2013}. We
refer to this quantity as the bar strength.

\begin{figure}
\begin{center}
\includegraphics[scale=0.525]{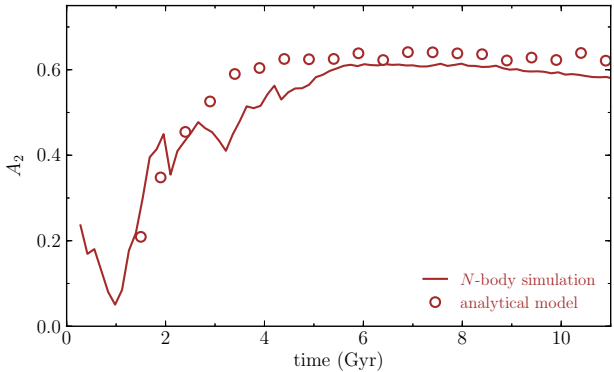}
\caption{(Colour online) Bar strength, measured from the simulation (line) and
from the analytical model (points). $A_2$ is the maximum relative contribution
of the $m=2$ Fourier component of the  mass distribution in the disc.}
\label{fig:A2}
\end{center}
\end{figure}

For the analytical model, we proceed in a way that allows us to treat it as if
it could be represented by particles. We extract from the simulation a random
sample of 100~000 initial conditions (i.e. positions and velocities of disc
particles) at a time $t_0=1.4$~Gyr (see Sect.~\ref{TDmodel}, where we use this
ensemble of orbits to further study dynamical trends). The orbits of each of
these `test particles' are then evolved forward in time in the presence of our
time-dependent analytical potential. Their successive positions can be treated
as if they were simulation particles. By stacking them at each time step, we
produce the snapshots in the bottom row of Fig.~\ref{fig:frames}. These mock
snapshots display a striking resemblance to the $N$-body snapshots, specially
bearing in mind that they were obtained by very indirect means. While this
comparison cannot be expected to yield a perfect morphological equivalence, one
notices that the bar lengths are in quite good agreement, and that in both
cases rings are present (although not of the same extent). The point is that
the dynamics that arises from the analytical model will give rise to very
similar disc and bar morphologies. In fact, the relative importance of the bar
is also quite comparable, as indicated by the $A_2$ parameter. Analogously to
the $N$-body case, we compute the $A_2$ of these mock snapshots and compare
them in Fig.~\ref{fig:A2}.

We must stress here that this comparison is an \textit{a posteriori}
verification, i.e. the bar strength of the $N$-body simulation was in fact not
used as an input to the analytical model. The fact the $A_2$ do agree well
counts as a further sign of the consistency of the constructed analytical
model.

It is clear, of course, that the variation of the bar strength modifies the
values of several parameters and yields richer information about the dynamics
of a self-consistent model. $N$-body simulations show that in general,
variations of the bar mass also change the mass ratios of the model's
components, the bar shape and the pattern speed of the galaxy. Hence, if one
wishes to use a mean field potential to `mimic' a self-consistent model as
accurately as possible, one should allow for all the parameters that describe
the bar (together with all other axisymmetric components) to depend on time,
assuming that the laws of such dependence were explicitly known. In our case,
however, we adopt a simpler approach and vary only the masses of the bar and
the disc, as a first step towards investigating such models when time-dependent
parameters are taken into account. Thus, we do not pretend to be able to
reproduce the exact dynamical evolution of a realistic galactic simulation.
Rather, we wish to understand the effects of time dependence on the general
features of barred galaxy models and compare the efficiency of chaos indicators
like the Generalized ALignment Index (GALI) method and the Maximal Lyapunov
Exponent (MLE) in helping us unravel the secrets of the dynamics in such
problems.

We stress that the method we introduced to construct the analytical model does
not rely -- at all -- on frozen potentials. Instead, it is grounded on the
detailed features of a  fully time-dependent, self-consistent $N$-body
simulation.

Unless otherwise stated, the units of the analytical model are given as: 1 kpc
(length), 1000 km$\cdot$ sec$^{-1}$ (velocity), 1 Myr (time), $2 \times 10^{11}
M_{\bigodot}$ (mass) and km$\cdot$ sec$^{-1} \cdot$ kpc$^{-1}$ ($\Omega_{b}$)
while the parameter $G=1$. The total mass $M_{tot}=M_{B}(t)+M_{D}(t)+M_{H}$ is
set equal to $3 \times 10^{11} M_{\bigodot}$ and since the halo's mass $M_H$ is
kept constant, the disc's mass $M_D(t)$ is varied as
$M_D(t)=M_{tot}-(M_H+M_B(t))$.

\section{Chaos detection techniques}\label{tools}

Let us here, for the sake of completeness, briefly recall how the two
main chaos detection methods used throughout the paper, namely the GALI and the
MLE, are defined and calculated. Considering the following TD 3-d.o.f.
Hamiltonian function which determines the motion of a star in a 3 dimensional
rotating barred galaxy:
\begin{equation}\label{eq:Hamilton}
  H=\frac{1}{2} (p_{x}^{2}+p_{y}^{2}+p_{z}^{2})+ V(x,y,z,t) -
  \Omega_{b}(t) (xp_{y}-yp_{x}).
\end{equation}
The bar rotates around its $z$--axis (short axis), while the $x$ direction is
along the major axis and the $y$ along the intermediate axis of the bar. The
$p_{x}$, $p_{y}$ and $p_{z}$ are the canonically conjugate momenta, $V$ is the
potential, $\Omega_b(t)$ represents the pattern speed of the bar and $H$ is the
total energy of the orbit in the rotating frame of reference (equal to the
Jacobi constant in the TI case).

The corresponding equations of motion are:
\begin{equation}\label{eq_motion}
\begin{array}{lcl}
  \dot{x} &=& \displaystyle p_{x} + \Omega_{b}(t) y, \\
  \dot{y} &=& \displaystyle p_{y} - \Omega_{b}(t) x,  \\
  \dot{z} &=& \displaystyle p_{z}, \\
  \dot{p_{x}} &=& \displaystyle -\frac{\partial V}{\partial x} + \Omega_{b}(t) p_{y}, \\
  \dot{p_{y}} &=& \displaystyle -\frac{\partial V}{\partial y} - \Omega_{b}(t) p_{x}, \\
  \dot{p_{z}} & =& \displaystyle -\frac{\partial V}{\partial z}, \\
\end{array}
\end{equation}
while the equations governing the evolution of a deviation vector
$\mathbf{w}=(\delta x,\delta y,\delta z,\delta p_{x},\delta p_{y},\delta
p_{z})$ needed for the calculation of the MLE and the GALI, are given by the
variational equations:
\begin{equation}\label{eq_dev_vect}
\begin{array}{lcl}
  \dot{\delta x} &=& \displaystyle \delta p_{x} + \Omega_b(t) \delta y,  \\
  \dot{\delta y} &=& \displaystyle \delta p_{y} + \Omega_b(t) \delta x,  \\
  \dot{\delta z} &=& \displaystyle\delta p_{z},\\
  \dot{\delta p_{x}} &=& \displaystyle- \frac{\partial^2 V}{\partial x \partial x}\delta x -
  \frac{\partial^2 V}{\partial x \partial y}\delta
  y - \frac{\partial^2 V}{\partial x \partial z} \delta z + \Omega_{b}(t) \delta p_{y}, \\
  \dot{\delta p_{y}} &=& \displaystyle- \frac{\partial^2 V}{\partial y \partial x}\delta x -
  \frac{\partial^2 V}{\partial y \partial y}\delta
  y - \frac{\partial^2 V}{\partial y \partial z} \delta z - \Omega_{b}(t) \delta p_{x}, \\
  \dot{\delta p_{z}} &=& \displaystyle- \frac{\partial^2 V}{\partial z \partial x}\delta x -
  \frac{\partial^2 V}{\partial z \partial y}\delta y - \frac{\partial^2 V}{\partial z \partial z} \delta z. \\
\end{array}
\end{equation}

Regarding the estimation of the value of the MLE, $\lambda_1$, of an orbit
under study we follow numerically its evolution in time together with its
deviation vectors $\mathbf{w}$, by solving the set of equations
(\ref{eq_motion}) and (\ref{eq_dev_vect}) respectively. For this task we use a
Runge-Kutta method of order 4 with a sufficiently small time step, which
guarantees the accuracy of our computations, ensuring the relative errors of
the Hamiltonian function (in the TI case) are typically smaller than $10^{-6}$.
Furthermore, we need to have a fixed time step in order to ensure that in the
TD case the orbits vary simultaneously with the potential.

In general, the derivatives of the potential $V$ depend explicitly on time and
and the ordinary differential equations (ODEs) (\ref{eq_motion}) are
non-autonomous. Hence, one has to solve together the equations for the
deviation vectors (\ref{eq_dev_vect}) with the equations of motion
(\ref{eq_motion}). Transforming the Eqs.~(\ref{eq_motion}) [and consequently
(\ref{eq_dev_vect})] to an equivalent autonomous system of ODEs by considering
time $t$ as an additional coordinate (see e.g.~\citealt[section
1.2b]{LichLieb}), is not particularly helpful, and is better to be avoided as
shown in \cite{GrySzl1995APPB}.

So, in order to compute the MLE and the GALI we numerically solve the
time-dependent set of ODEs (\ref{eq_motion}) and (\ref{eq_dev_vect}). Then,
according to \cite{BenGalStr1976PRA,ConGalGio1978PRA,Ben1980Mecc} the MLE
$\lambda_1$ is defined as:
\begin{equation}
\label{LE}
\lambda_1 =\lim_{t \rightarrow \infty} \sigma_1(t),
\end{equation}
where:
\begin{equation}
\label{sigma_1}
\sigma_1(t)= \frac{1}{t}
\ln \frac{\|\mathbf{w}(t)\|}{\|\mathbf{w}(0)\|},
\end{equation}
is the so-called `finite time MLE', with $\|\mathbf{w}(0)\|$ and
$\|\mathbf{w}(t)\|$ being the Euclidean norm of the deviation vector at times
$t=0$ and $t>0$ respectively. A detailed description of the numerical algorithm
used for the evaluation of the MLE can be found in \cite{SkoLNP2010}.

This computation can be used to distinguish between regular and chaotic orbits,
since $\sigma_1(t)$ tends to zero (following a power law $\propto t^{-1}$) in
the former case, and converges to a positive value in the latter. But
Hamiltonian (Eq.~\ref{eq:Hamilton}) is TD, which means that its orbits could change
their dynamical behavior from regular to chaotic and vice versa, over different
time intervals of their evolution. In such cases, the MLE (\ref{LE}) does not
behave exactly as in TI model (presenting in general stronger fluctuations) and
its computation might not be able to identify the various dynamical phases of
the orbits, since by definition it characterizes the asymptotic behavior of an
orbit \citep[see e.g.][]{ManBouSkoJPhA2013}. Nevertheless, we will show the MLE
for a number of orbits throughout the paper, for a more global discussion of
the several dynamical properties observed.

Thus, in order to avoid such problems in our study, we use the GALI method of
chaos detection \citep{SkoBouAntPhyD2007}. The GALI index of order $k$
(GALI$_k$) is determined through the evolution of $2 \leq k \leq N$ initially
linearly independent deviation vectors $\textbf{w}_i(0)$, $i = 1,2,\ldots,k$,
with $N$ denoting the dimensionality of the phase space of our system. Thus,
apart from solving Eqs.~(\ref{eq_motion}), which determines the evolution of an
orbit, we have to simultaneously solve Eqs.~(\ref{eq_dev_vect}) for each one of
the $k$ deviation vectors. Then, according to \cite{SkoBouAntPhyD2007},
GALI$_k$ is defined as the volume of the $k$-parallelogram having as edges the
$k$ unit deviation vectors
$\hat{\textbf{w}}_i(t)=\textbf{w}_i(t)/\|\textbf{w}_i(t)\|$, $i = 1,2,...,k$.
It can be shown, that this volume is equal to the norm of the wedge product
(denoted by $\wedge$) of these vectors:
\begin{equation}\label{GALI:0}
  \textrm{GALI}_{k}(t)=\parallel \hat{\textbf{w}}_{1}(t) \wedge \hat{\textbf{w}}_{2}(t) \wedge \ldots \wedge \hat{\textbf{w}}_{p}(t) \parallel.
\end{equation}
We note that in the above equation the $k$ deviation vectors are normalized but
their directions are kept intact.

In principal and for TI systems, the GALI$_k(t)$ for regular orbits remains
practically constant and positive if $k$ is smaller or equal to the
dimensionality of the torus on which the motion occurs, otherwise, it decreases
to zero following a power law decay. For the chaotic ones, all GALI$_k(t)$ tend
exponentially to zero with exponents that depend on the first $k$ LEs of the
orbit \citep{SkoBouAntPhyD2007,SkoBouAntEPJST2008}. Nevertheless, in the TD
case studied in \cite{ManBouSkoJPhA2013} and also here, the way the theoretical
estimation of the GALI's exponential rates are strongly related to the LEs,
being more complicated and still open to further inquiry.

The procedure used for the detection of the several different dynamical epochs
of the TD system is the following: We evolve the GALI$_k$ with $k=2$ or $k=3$
(i.e., using 2 or 3 deviation vectors) for the 2-d.o.f or 3-d.o.f. cases
respectively and whenever GALI$_k$ reaches very small values (i.e.~GALI$_k \leq
10^{-8}$) we re-initialize its computation by taking again $k$ new random
orthonormal deviation vectors, which resets the GALI$_k=1$. We allow then these
vectors to evolve under the current dynamics. For time intervals where the
index decays exponentially corresponds to chaotic epochs while in the other
cases to non-chaotic. The reason in doing this is that we need to follow the
current dynamical stability of an orbit under-study which in principle can
interplay between chaotic and regular for different epochs during the total
time evolution. Thus, let us assume that a trajectory experiences chaotic
dynamics and later on drifts to a regular regime. The volume formed by the
deviation vectors will first shrink exponentially to very small values and
remain small throughout the whole evolution unless one re-initializes the
deviation vectors and their volume, in order to allow them to `feel' the new
current dynamics. However, when we are interested in more general dynamical
trends in time (less details), we will fix time intervals and we will
re-initialize the deviation vectors in the beginning of each one.

\section{The 2-d.o.f. time-independent case} \label{2DTImodel}

Shedding some light on the underlying TI dynamics is an important step for
understanding the more complicated case of the fully 3-d.o.f. TD model where
all parameters vary simultaneously in time.

By setting $z,p_z$ equal to zero at $t=0$ (remaining zero at all times) in the
Hamiltonian Eq.~(\ref{eq:Hamilton}), the orbits' motion is then restricted in
the 2 dimensional $(x,y)$ space. Note that here $t=0$ refers to the
$t_0=1.4$~Gyr of the $N$-body simulation. We can then study, in a frozen
potential, individual orbits and the stability of the phase space, in terms of
detecting and locating chaotic and regular motion, for the several sets of the
potential parameters at deferent times, as derived from the $N$-body
simulation. We shall begin by choosing fixed parameters from four time
snapshots, i.e., at $t=1.4$~Gyr, $t=4.2$~Gyr, $t=7.0$~Gyr and $t=11.2$~Gyr and
we integrate orbits for 10~Gyr.
\begin{figure*}
\begin{center}
\includegraphics[scale=1]{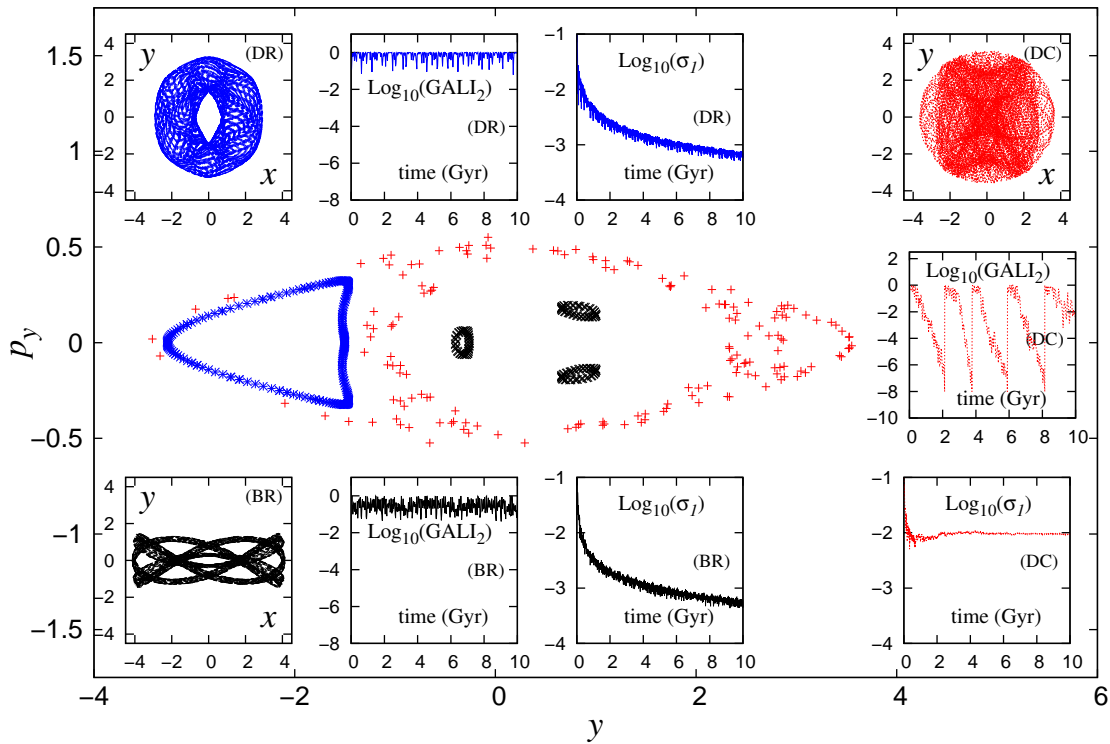}
\caption{(Colour online) The Poincar\'e Surface of Section defined by $x=0$,
$p_x \geq 0$ with $H=-0.19$, for three typical orbits (two regular and one
chaotic) being integrated for 10~Gyr. The set of parameters for the bar, disc
and halo components are chosen from the fits with the 3-d.o.f. TD Hamiltonian
at $t=7.0$~Gyr of the $N$-body simulation. In the insets we depict their
projection on the $(x,y)$-plane together with the GALI$_2$ and MLE $\sigma_1$
evolution in time (see \protect{Table~\ref{tab1}} for the exact parameters and
text for more details on these trajectories).} \label{fig2Dpss}
\end{center}
\end{figure*}
In Fig.~\ref{fig2Dpss}, we present the Poincar\'e Surface of Section (PSS)
defined by $x=0$, $p_x \geq 0$ with $H=-0.19$, for three typical orbits being
integrated for 10~Gyr. The set of parameters for the bar, disc and halo
components are chosen from the fits with the 3-d.o.f. TD Hamiltonian at
$t=7.0$~Gyr of the $N$-body simulation (see Table~\ref{tab1} for more details).
The blue $(\ast)$ points on the PSS correspond to a disc regular orbit, forming
a curve by the successive intersections with the plane $x=0$, having initial
condition $(y,p_y)=(-1.5,0.0)$ with $x=0$ and $p_x=H(x,y,p_y)$ (called `DR'
from now on). Its projection on the $(x,y)$-plane is shown in the top left
inset panels of Fig.~\ref{fig2Dpss} and coloured in blue. The GALI$_2$ for this
orbit confirms that its motion is regular by oscillating to a positive value
during its evolution in time as well as the MLE $\sigma_1$ following a power
law decay [see second and third top inset panels of Fig.~\ref{fig2Dpss}
respectively (blue), note that the axis here are in lin-log scale]. The three
small black curves in the central part of Fig.~\ref{fig2Dpss}, marked with
($\times$), are formed by the successive intersections of an initial condition
with $(y,p_y)=(-0.4,0.0)$ (we will call it `BR'). From its projection on the
$(x,y)$-plane [first bottom inset panel of Fig.~\ref{fig2Dpss} (black)], it is
evident that it is a bar-like orbit elongated along the long $x$-axis. It
surrounds a stable periodic orbit of period 3 in the center of these islands
and its regular dynamics is clearly revealed from the evolution of the GALI$_2$
and the MLE $\sigma_1$ evolution [second and third top (black) inset panels of
Fig.~\ref{fig2Dpss} respectively, note that the axis are again in lin-log
scale]. The central scattered red points on the PSS, marked with $(+)$,
correspond to a chaotic orbit with initial condition $(y,p_y)=(2.5,0.0)$
(called `DC' from now on). In the three inset panels positioned vertically in
the right part of the Fig.~\ref{fig2Dpss} (red), we depict its projection on
the $(x,y)$-plane (top panel). Its GALI$_2$ successive and exponential decrease
to zero in time (middle panel) indicates its chaotic nature. Notice that we
re-initialize the deviation vectors each time the index becomes small $(\leq
10^{-8})$. Its MLE $\sigma_1$, as expected, converges to a positive value
(bottom inset).

\begin{figure}
\begin{center}
\includegraphics[scale=0.6]{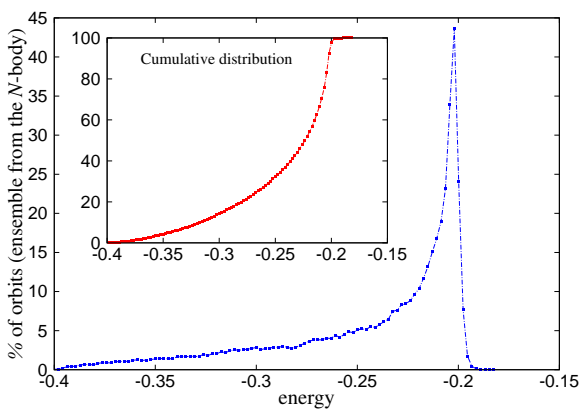}
\caption{(Colour online) Dispersion energy histogram for an ensemble of 100~000
initial conditions from the $N$-body simulation. The energies are calculated
from the TD Hamiltonian function Eq.~(\ref{eq:Hamilton}) at $t_0=1.4$~Gyr. In
the inset panel we show the cumulative distribution of this set of orbits.}
\label{fig:NbodyEN}
\end{center}
\end{figure}

Since in this case, the potential has no TD parameters, the general asymptotic
nature can be either regular or chaotic (weakly or strongly). This of course
will change when the parameters start to vary in time and the motion can
convert from one kind to the other, though this will be driven by the momentary
underlying dynamics. Hence, it is important to have a good idea of how the
phase space (through the potential parameters) evolves in time and for
different values of the Hamiltonian function. Regarding the former we will pick
four sets of parameters given at different times as extracted from the $N$-body
simulation, i.e., at $t=1.4$~Gyr when the bar is still relatively small, at the
end of the simulation ($t=11.2$~Gyr) and at two intermediate states
($t=4.2$~Gyr and $t=7.0$~Gyr). The next step is to pick some `representative'
or `most significant' energy (Hamiltonian function) values from the total
available energy spectrum at each time and set of parameters that would be
relevant for the $N$-body simulation as well. For this reason, we first
calculated the energy dispersion of the 100~000 disc particles from the
simulation. The energies are calculated from the TD Hamiltonian function
Eq.~(\ref{eq:Hamilton}) at $t_0=1.4$~Gyr. In the inset panel we show the
cumulative distribution of this set of orbits. From the energy distribution
histogram in Fig.~\ref{fig:NbodyEN},
we can see that the majority of the trajectories of the $N$-body simulation are
concentrated in relatively large values.
\begin{figure*}
\begin{center}
\includegraphics[scale=0.395]{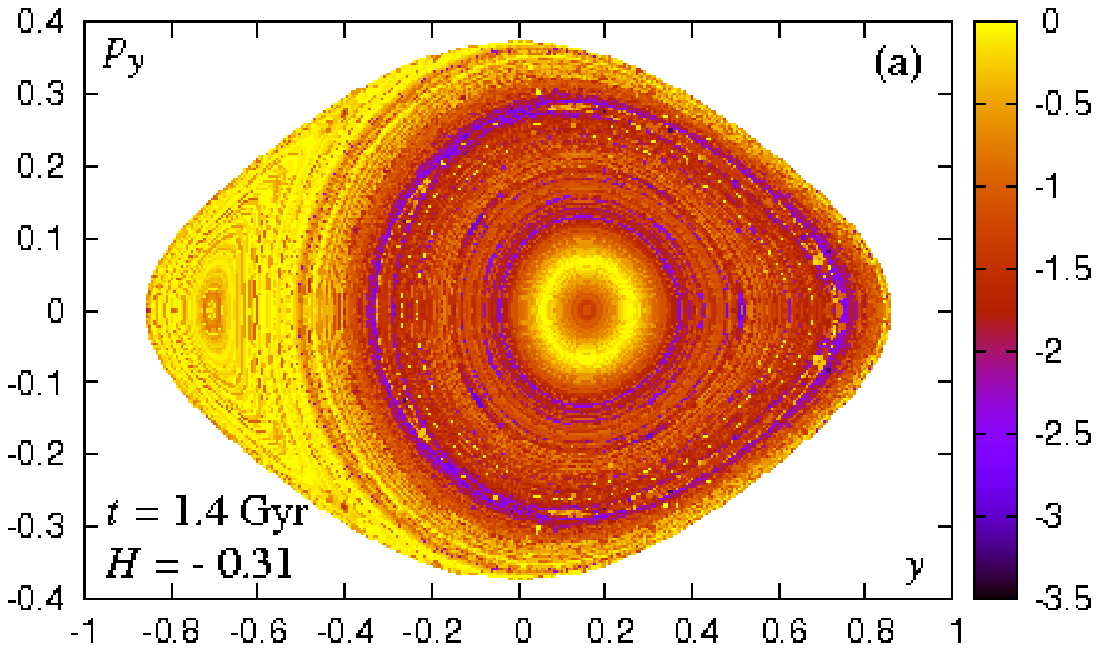}
\includegraphics[scale=0.395]{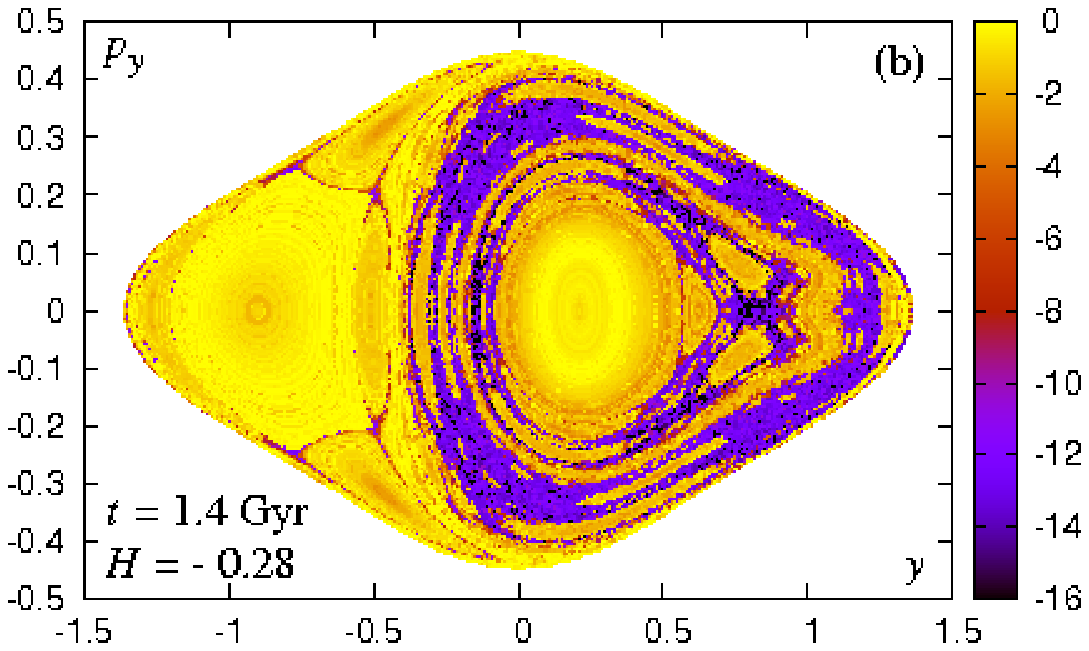}
\includegraphics[scale=0.395]{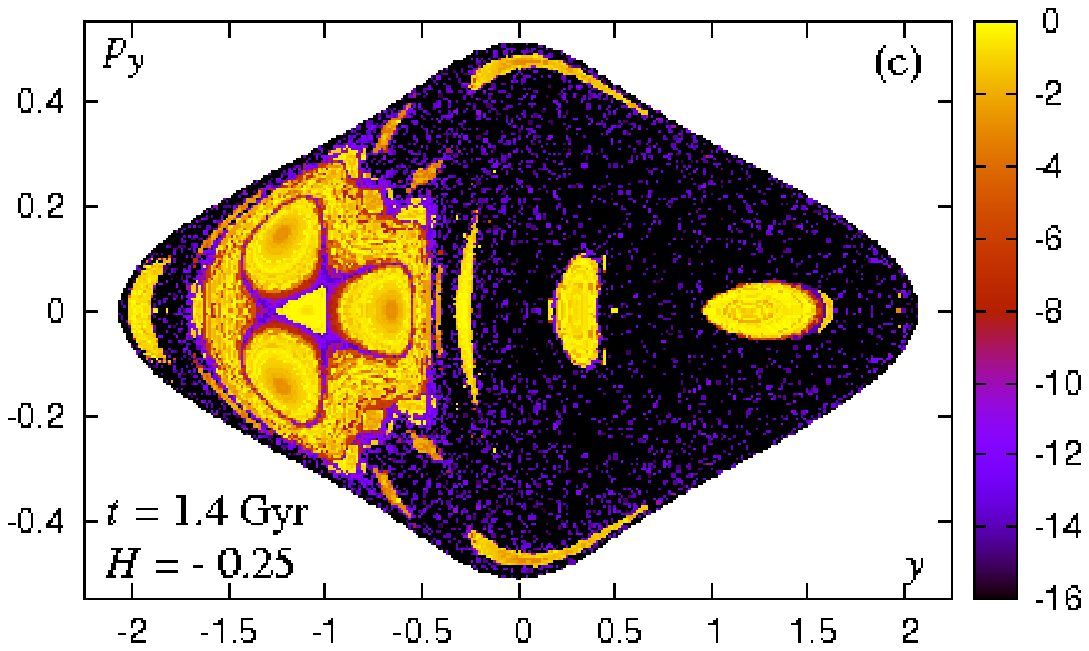}
\includegraphics[scale=0.395]{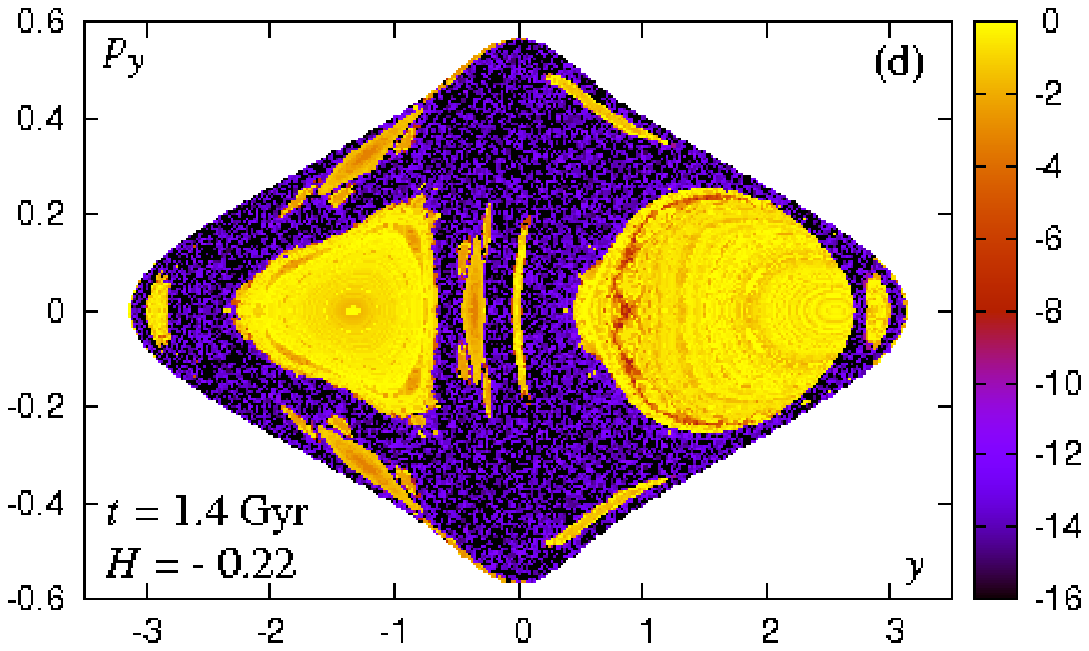}
\includegraphics[scale=0.395]{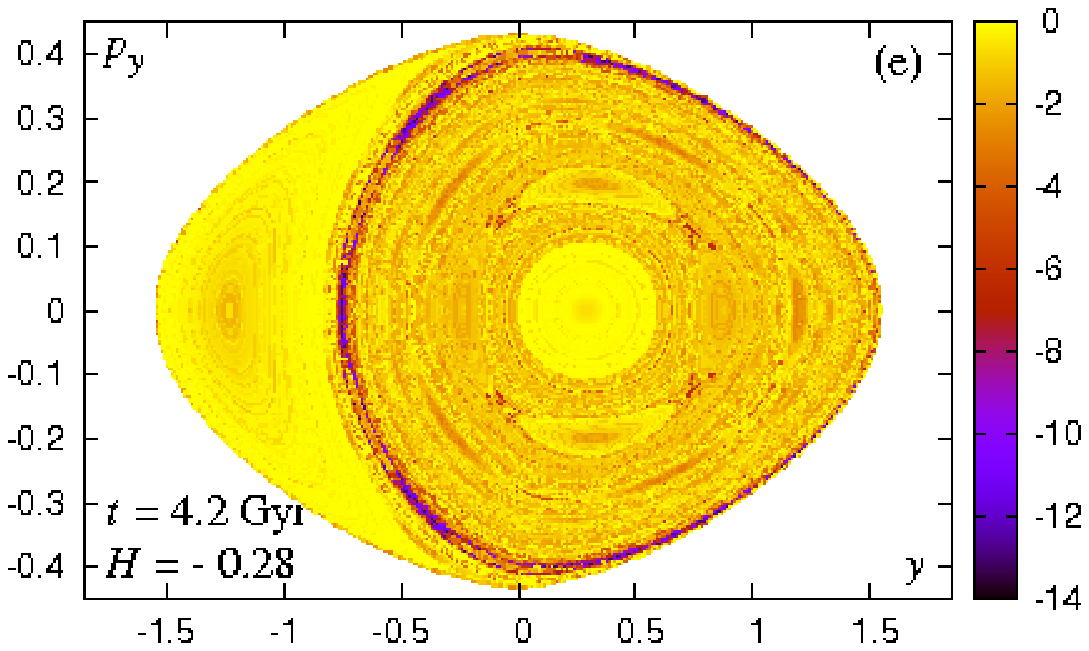}
\includegraphics[scale=0.395]{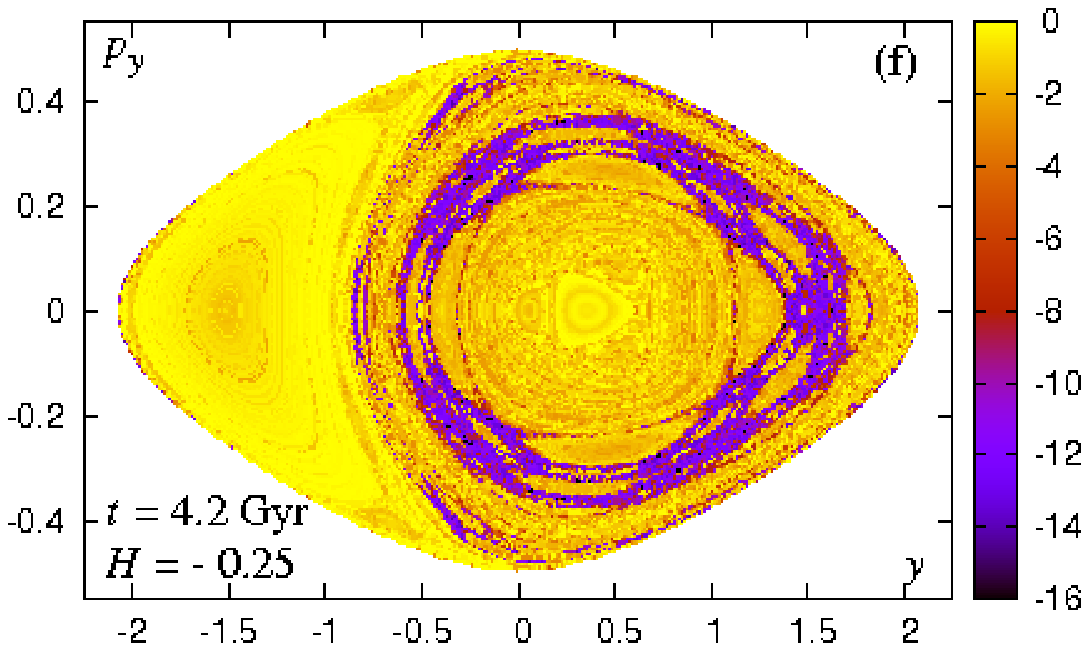}
\includegraphics[scale=0.395]{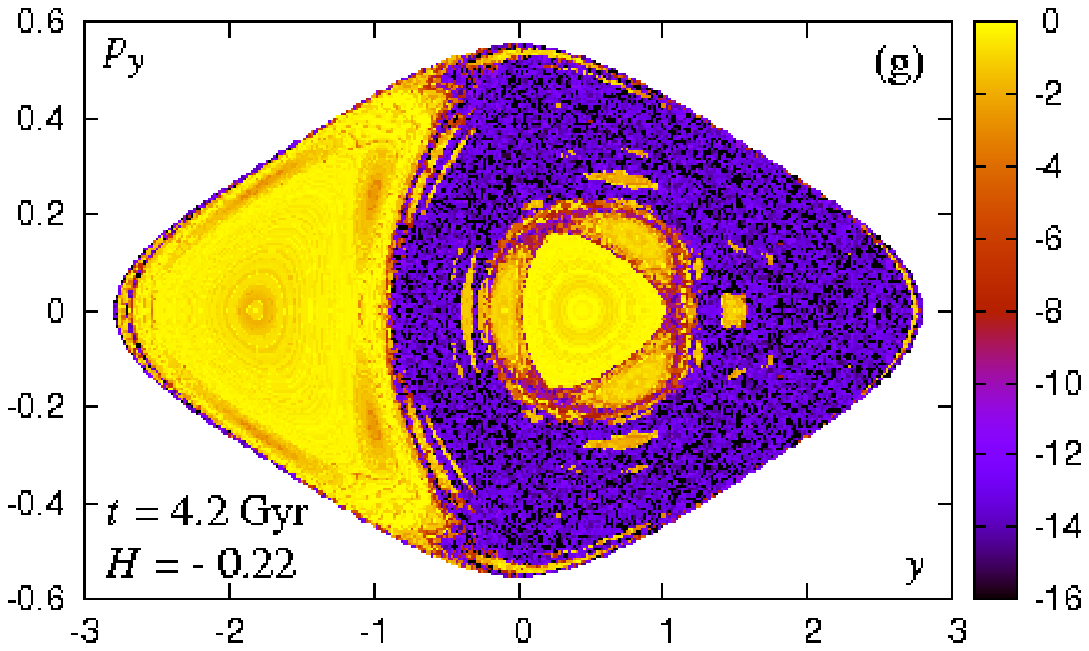}
\includegraphics[scale=0.395]{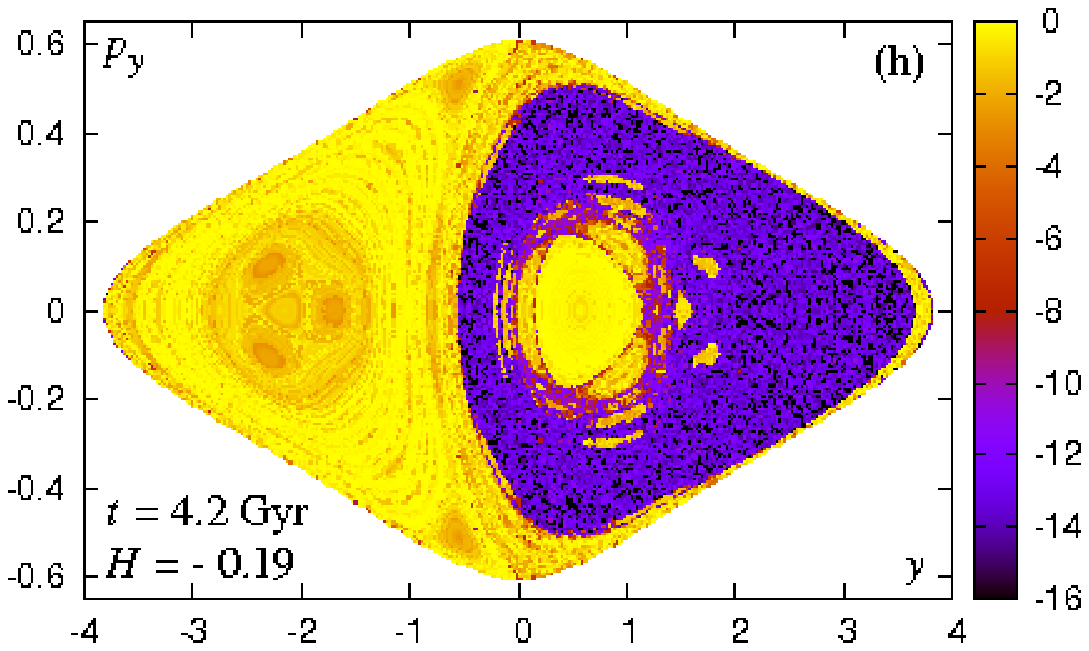}
\includegraphics[scale=0.395]{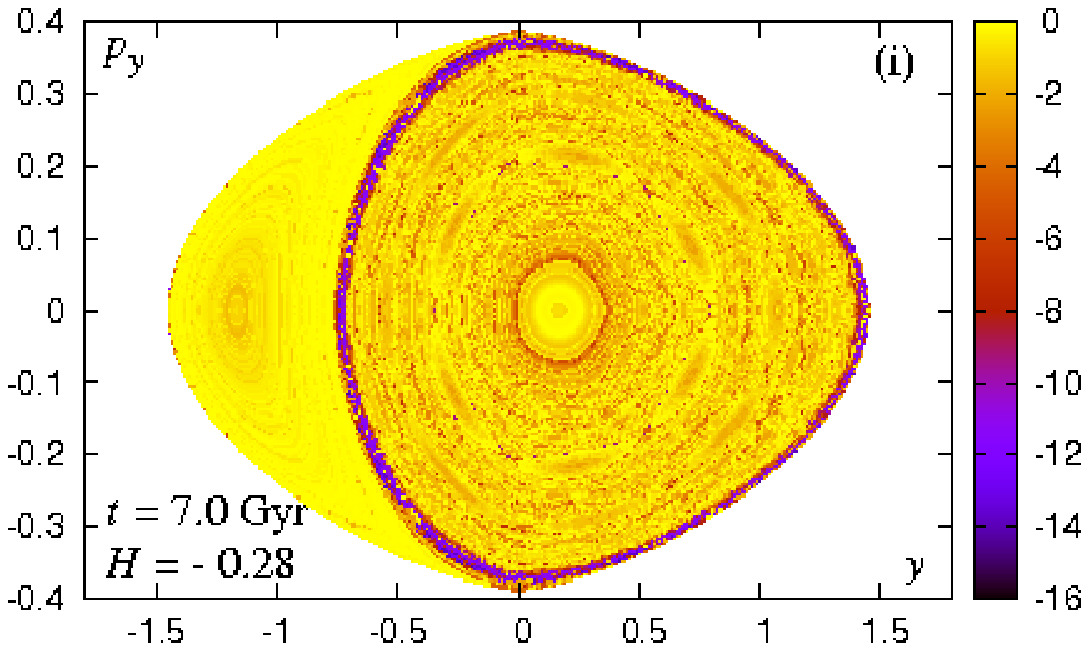}
\includegraphics[scale=0.395]{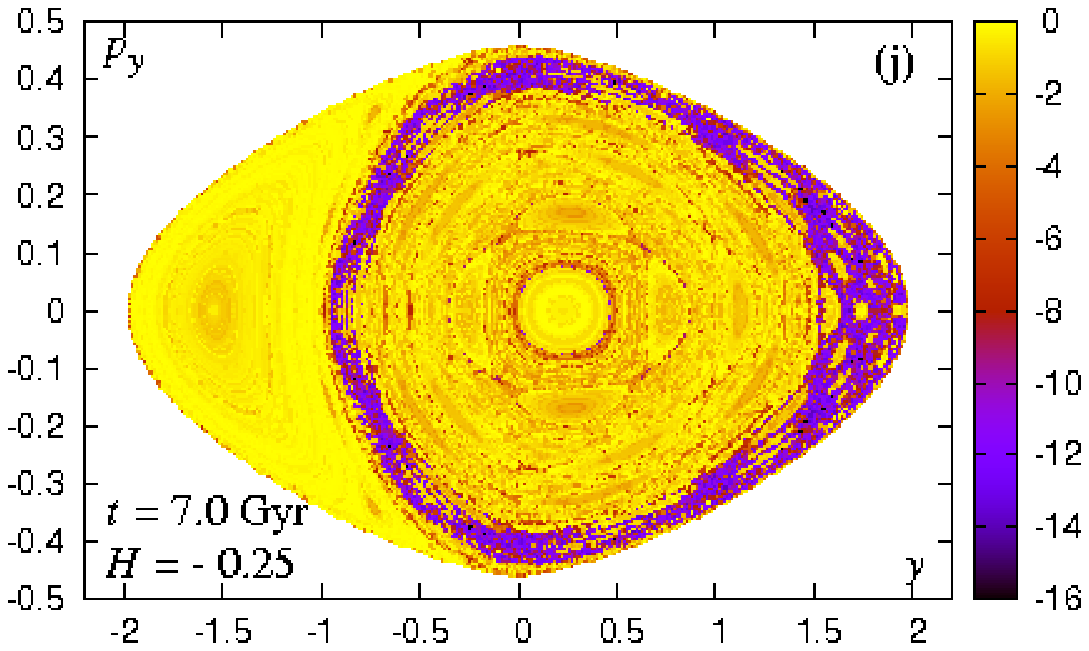}
\includegraphics[scale=0.395]{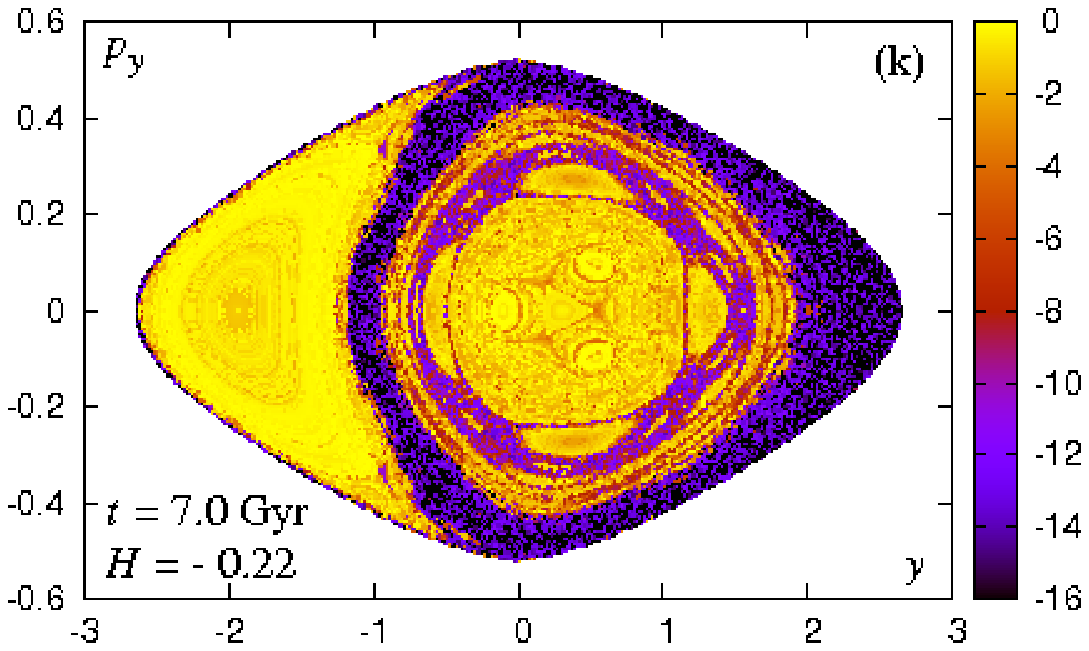}
\includegraphics[scale=0.395]{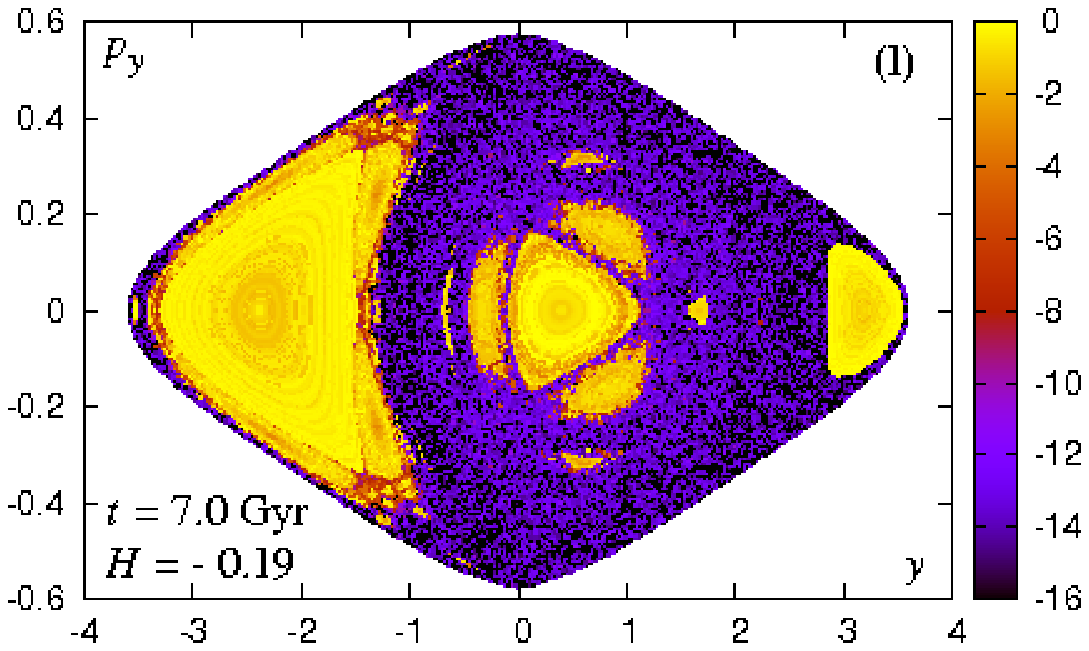}
\includegraphics[scale=0.395]{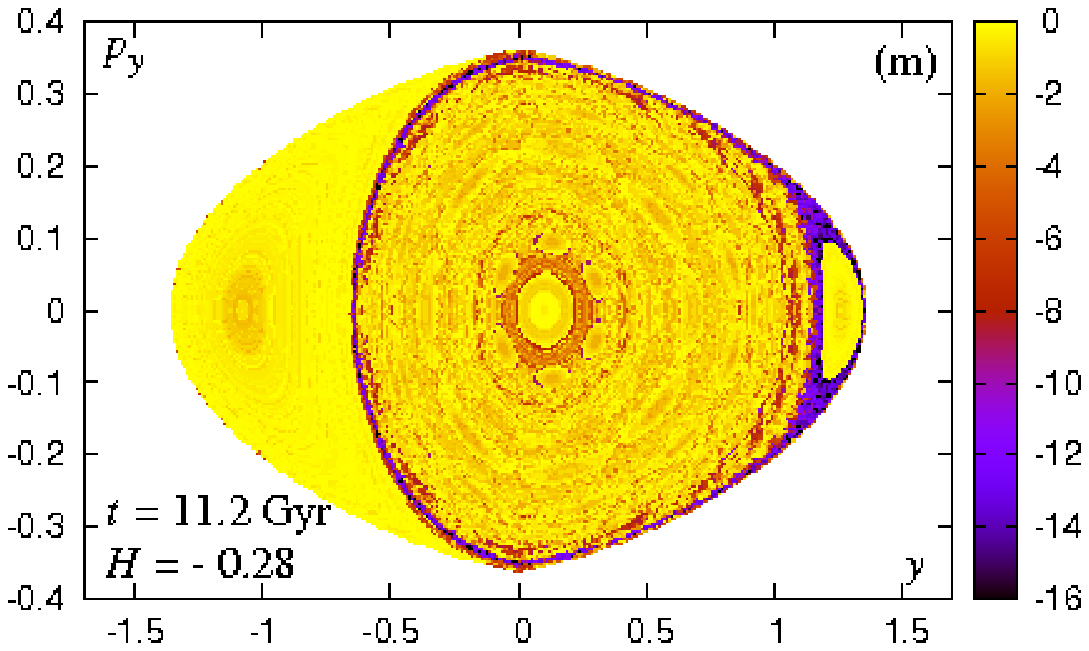}
\includegraphics[scale=0.395]{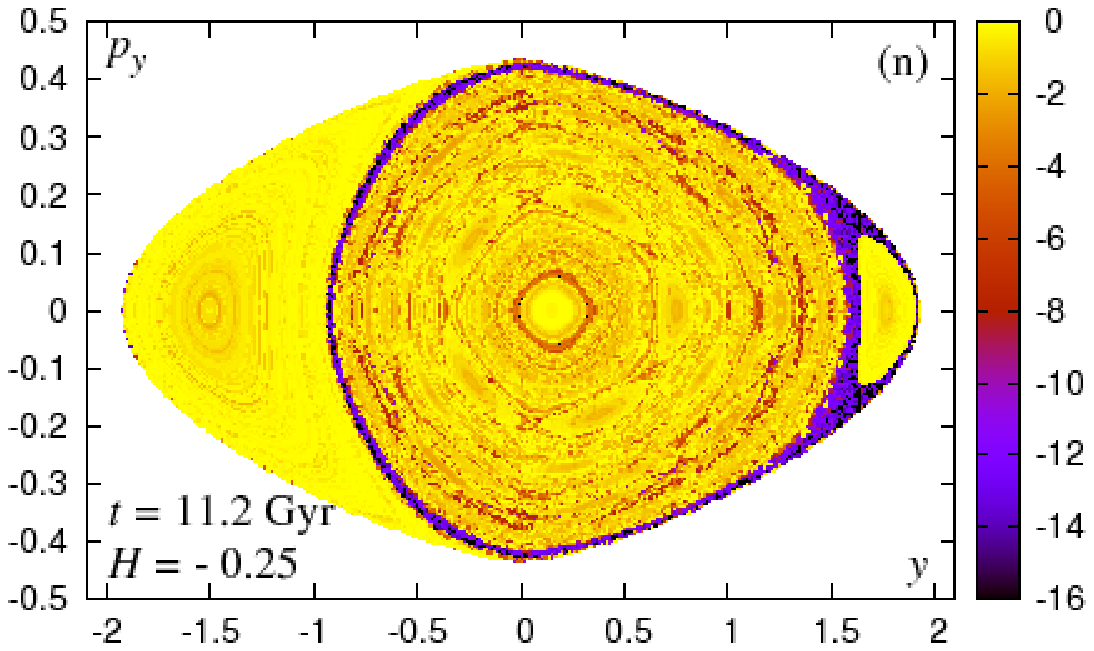}
\includegraphics[scale=0.395]{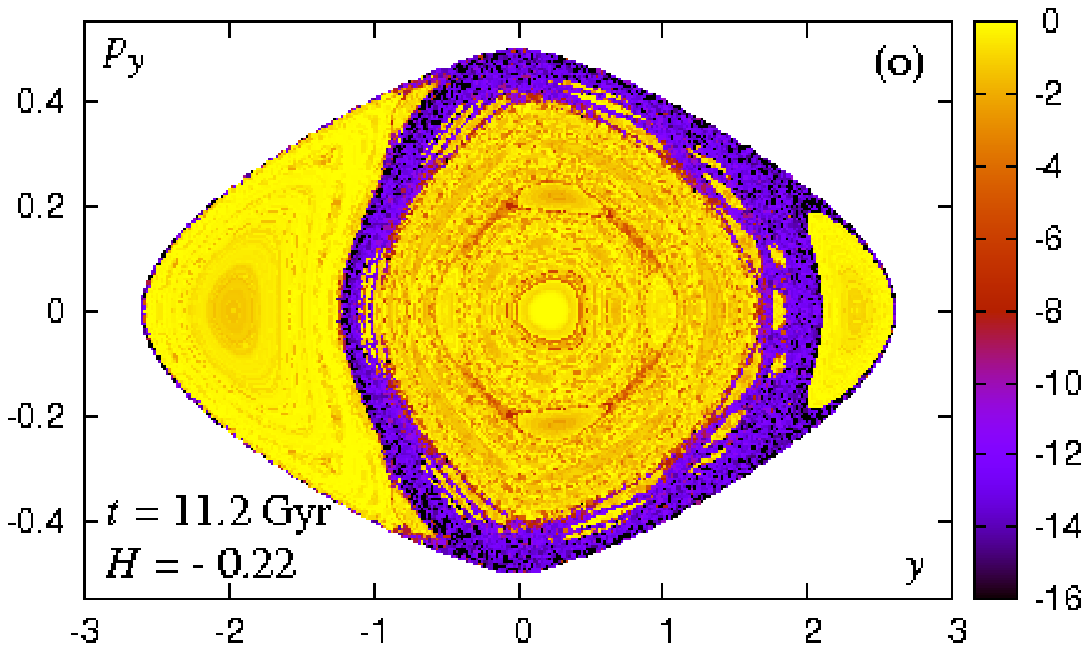}
\includegraphics[scale=0.395]{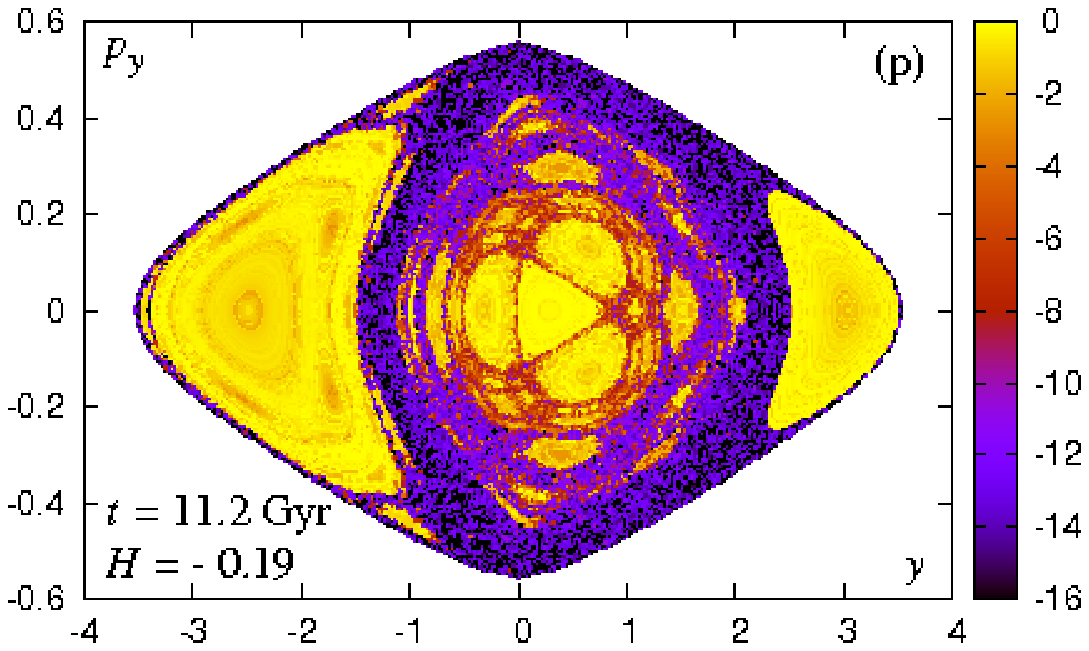}
\caption{(Colour online) The (in)stability map for the 2-d.o.f. frozen
potential case using a grid of $\approx$ 100~000 initial conditions on the PSS
and integrating them for 10~Gyr. The colour-bar represents the final GALI$_2$
values of each initial condition in the end of the iteration. The first row
refers to a set of potential parameter given at $t=1.4$~Gyr for Hamiltonian
values $H=-0.31, -0.28, -0.25, -0.22$, the second row at $t=4.2$~Gyr for
$H=-0.28, -0.25, -0.22, -0.19$, the third row at $t=7.0$~Gyr for $H=-0.28,
-0.25, -0.22, -0.19$ and the fourth row at $t=11.2$~Gyr for $H=-0.28, -0.25,
-0.22, -0.19$. The yellow (light-grey in b/w) colour corresponds to regular
orbits where the GALI$_2$ oscillates around to relatively large positive
values, the black color represents the chaotic orbits where GALI$_2$ tends
exponentially small values, while the intermediate colors in the colour-bars
between the two represent `weakly chaotic or sticky' orbits. The exact set of
parameters used at each $t=1.4,4.2,7.0,11.2$~Gyr are given in
table~\ref{tab1}.} \label{fig:2Dgalipss}
\end{center}
\end{figure*}

Exploiting this information, we first choose a sample of 2-d.o.f. Hamiltonian
function values (for the four times mentioned above and the respective sets of
parameters), from the interval of energies where the majority of the $N$-body
simulation's particles is more probable to be found. From this sample, we
select a subset of representative energies to illustrate typical phase space
structures, focusing at this point on the underlying dynamics. Then, we chart
the regular and chaotic regimes of the phase space with GALI$_2$. In
Fig.~\ref{fig:2Dgalipss}, we have used a grid of 100~000 initial conditions on
the $(y,p_{y})$-plane of the corresponding PSS and we have constructed a
(colour online) chart of the chaotic and regular regions similar to the PSS,
but with more accuracy and higher resolution, in a similar manner just like in
\cite{ManAthMNRAS2011}, using the GALI$_2$ method. The different colour
corresponds to the different final value of the GALI$_2$ after 10~Gyr (10~000
time units) for orbits representing each cell of the grid. The yellow
(light-grey in b/w) colour corresponds to regular orbits (and areas) where the
GALI$_2$ oscillates around to relatively large positive values, the black color
represents the chaotic orbits where GALI$_2$ tends exponentially to zero
(10$^{-16}$), while the intermediate colors in the colour-bars between the two
represent `weakly chaotic or sticky' orbits, i.e. orbits that `stick' onto
quasi-periodic tori for long times but their nature is eventually revealed to
be chaotic. Note that the model in this case is TI and hence there is no need
for re-initialization of the deviation vectors, since the asymptotic dynamical
nature of the orbits does not change in time. The first row refers to a set of
potential parameter given at $t=1.4$~Gyr for Hamiltonian values $H=-0.31,
-0.28, -0.25, -0.22$, the second row at $t=4.2$~Gyr for $H=-0.28, -0.25, -0.22,
-0.19$, the third row at $t=7.0$~Gyr for $H=-0.28, -0.25, -0.22, -0.19$ and the
fourth row at $t=11.2$~Gyr for $H=-0.28, -0.25, -0.22, -0.19$.

\begin{figure}
\begin{center}
\includegraphics[scale=0.6]{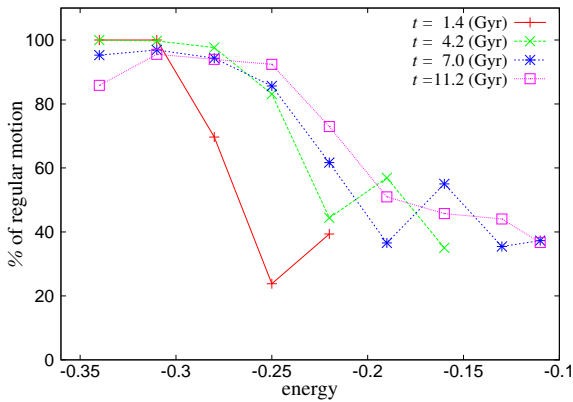}
\caption{(Colour online) Percentages of regular motion for the 2-d.o.f. frozen
case as a function of the energy $H$ for the different sets of parameters at
$t=1.4$~Gyr, $t=4.2$~Gyr, $t=7.0$~Gyr and $t=11.2$~Gyr.}
\label{fig:2DOFfroperc}
\end{center}
\end{figure}

Using the above approach, we can measure and quantify the variation of the
percentage of regular orbits in the phase space as the total energy increases
for a specific choice of potential parameters at same fixed times. The chosen
values of the Hamiltonian functions cover the range of the available energy
interval up to the value of the escape energy which in general is different.
Although the main general trend is that this percentage decreases as the energy
grows, its behavior changes at high energy values where it is no more
monotonic. Note that this happens for energy values $H>-0.19$ out of the range
of $N$-body simulation orbits. In Fig.~\ref{fig:2DOFfroperc} we show the
variation of percentages of regular motion as a function of the energy $H$ for
the different sets of parameters at $t=1.4$~Gyr, $t=4.2$~Gyr, $t=7.0$~Gyr and
$t=11.2$~Gyr. The threshold GALI$_2 \geq 10^{-8}$ was used to characterize an
orbit as regular and GALI$_2 < 10^{-8}$ as chaotic which will also be the chaos
criterion/threshold throughout this paper. We should emphasize that these
percentages refer to a set of initial conditions that cover uniformly the whole
2-d.o.f. phase space. On the other hand, an ensemble of trajectories extracted
from the $N$-body simulation does not necessarily populate `democratically' the
phase space. By simply inspecting these percentages one can claim that the
fraction of regular motion is systematically larger for later times, and for
all energies. When looking and comparing the phase space for early times, i.e.,
$t=1.4, 4.2$~Gyr (first and second row) and late times, i.e., $t=7.0, 11.2$~Gyr
(third and fourth row) in Fig.~\ref{fig:2Dgalipss}, we may see that the central
island of stability, originating bar-like orbits, is becoming larger as the
time grows and this is even more evident for the relatively larger energies
(see the third and fourth row from top to bottom). This indicates that the bar
component becomes gradually more important and dominant.

Considering the $N$-body simulation's bar growth evolution
(Fig.~\ref{fig:frames}), this fact seems to be already in a quite good
agreement, since one would expect the barred morphological features to become
more evident as time increases and for the relevant energies, in both the
simulation and the derived TD analytical model. We may here note that the
energies of the $N$-body simulation particles (being 3 dimensional objects) are
concentrated in the interval $-0.3 \lesssim H \lesssim -0.2$ (as shown in
Fig.~\ref{fig:NbodyEN}). A more `realistic' ensemble of initial conditions will
be used for the 3-d.o.f. case in the next section. Despite the fact that these
percentages refer to the 2-d.o.f. case, one may get a brief idea of how the
relative stability of the phase space changes as the energy varies in the
`full' 3-d.o.f. model and expect a relatively large fraction of regular motion
for an ensemble chosen within this energy interval.

\section{The 3-d.o.f. time-dependent model} \label{TDmodel}

For the global study of the dynamics of the phase and configuration space of a
3-d.o.f. model (TI and/or TD), the choice of initial conditions plays a crucial
role. When one seeks to explore the whole available phase space the orbits
might occupy, a few useful approaches have been proposed and used in the
literature \citep[e.g.][]{Sch1993ApJ,PapLas1998A&A,ElZantShlosman2002}, where
some adequate (but different) ways in populating several different families of
orbits were proposed by giving initial conditions on several appropriate chosen
planes in positions and momenta with the total energy restriction taken into
account \citep[also used recently in][]{MafDarCinGio2013MNRAS}. In
\cite{ManAthMNRAS2011} and \cite{ManBouSkoJPhA2013} the authors used two
similar distributions as in the latter references and also a third one, derived
by a random set of orbits related to the density mass distribution of the
model. Their momenta were set zero along the $y,z$ directions while the $p_x$
was estimated by the total Hamiltonian function value, randomly chosen from the
available energy level. The approaches described above may supplementary
explore well the phase space of a model. However, they do not ensure that the
particles of a realistic $N$-Body simulation necessarily populate the phase
space in the same uniform way. This becomes even more complicated when the
potential under consideration ceases to be TI and the parameters'
time-dependencies in time cause consecutive alternations in the phase space.

For these reasons, here, and in order to study \textit{global stability and
dynamical trends} in our 3-d.o.f. TD analytical model, we extract an ensemble
of 100~000 initial conditions directly from the $N$-body simulation at the time
$t_0=1.4$~Gyr where the bar has already started to be formed and starts growing
from that point on. Note that these orbits are chosen only from the disc
density distribution of the simulation and do not include halo particles. Our
goal is twofold: (i) to estimate quantitatively the fraction of chaotic and
regular motion as time increases and (ii) to check whether and how this
variation is associated now to the bar strength for this TD model. From this
point and on, we attribute the time $t_0=1.4$~Gyr to $t_0=0$. Then, in the TD
analytical model, the orbits are integrated for $10$~Gyr (a bit less than a
Hubble time), a realistic upper limit related to the $N$-body simulation's set
up.

Let us point out here that the straightforward comparison of the two approaches
is rather difficult, since several assumptions have been made for the
construction of the TD analytical model. For example, the total energy of the
particles is not expected to be conserved, like in the $N$-body simulation
where the self-consistency ensures this with a good accuracy. Moreover, halo
particles, present in the $N$-body simulation, are not taken into account as
point mass particles at this point. The reason, as explained earlier, is that
we are mainly interested in the bar's growth and evolution in time. Hence, when
one allows all the TD model parameters of the several components of model to
vary in time for a set of initial conditions, the total energy is not
conserved. Nevertheless, and for the sake of a more general study, we tried two
alternative cases; let us call them \textit{dynamical scenarios} whose
parameters are given in such a way that the total energy can be conserved by
making reasonable assumptions. We then checked how sensitive the general
\textit{dynamical trends} are to these choices of potentials.

\begin{description}

\item[\textit{Scenario A}:] In this case, all the parameters of the three components
(bar-disc-halo) of the potential vary simultaneously in time, following the
fitting functions derived from the simulation. In this case the value of the
Hamiltonian in general is not conserved in time for a single trajectory. Here,
the TD model incorporates the time-dependencies of the $N$-body simulation for
each particle, however in the latter the sum of the energy of all the particles
remains approximately constant. This is the evolutionary scenario
presented in the panels of the lower row in Fig.~\ref{fig:frames}. We can see
that this TD model can capture quite successfully the basic trends and the bar
formation of the $N$-body simulation (upper row).\\

\item[\textit{Scenario B}:] In this case, the parameters of the bar and disc vary as
in case `Scenario A' but the halo ones are adjusted in such a way that the
value of the total Hamiltonian function, for each initial condition, remains
constant in time. In order to achieve that, we first allow the bar and disc
potential to vary in time as before, keeping constant the halo parameter
$\gamma$ ($\approx 0.234$ at $t_0=1.4$~Gyr) and then we allow the $a_H$ to vary
at each time step. From the initially known Hamiltonian value, we calculate the
value of the total potential at every time step and then find the value of the
$V_H(t)$ such as the total energy remains constant. Having this value, we can
numerically estimate the appropriate value of $a_H(t)$ that fulfills the
imposed energy conservation condition. This also implies that the scale radius
of the halo might vary in a similar way for all the orbits but not identically so.\\

\item[\textit{Scenario C}:] In this case, the bar parameters are completely time-dependent,
the disc parameters are fixed in an approximately averaged value, with respect
to the whole evolution, i.e.,  $A=0.8$ and $B=0.55$, while the halo parameters
are estimated just like in case `Scenario B', obtaining again the energy
conservation.

\end{description}

\begin{figure*}
\begin{flushleft} \qquad \qquad \qquad  \textit{Scenario A} (bar-like orbit $B1$) \end{flushleft}
\begin{center}\hspace{0.9cm}
\includegraphics[scale=0.475]{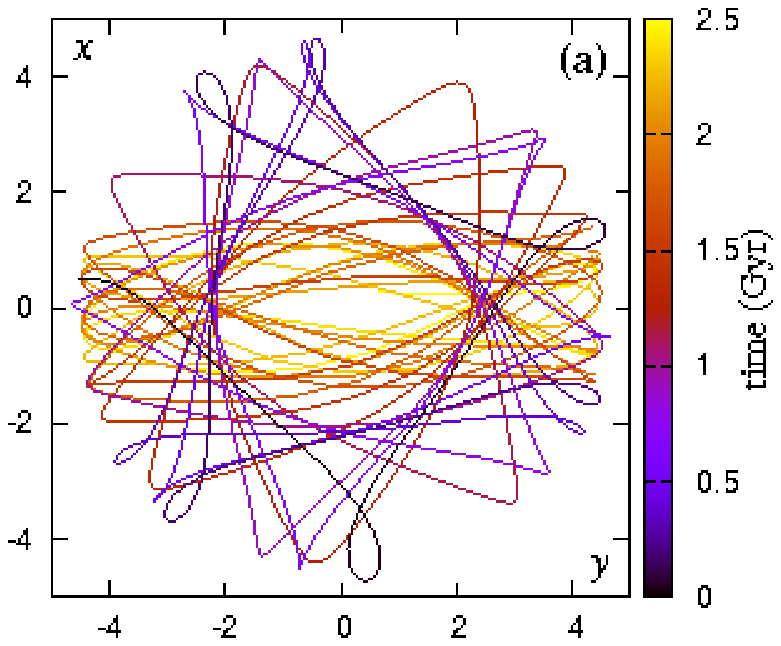}
\includegraphics[scale=0.475]{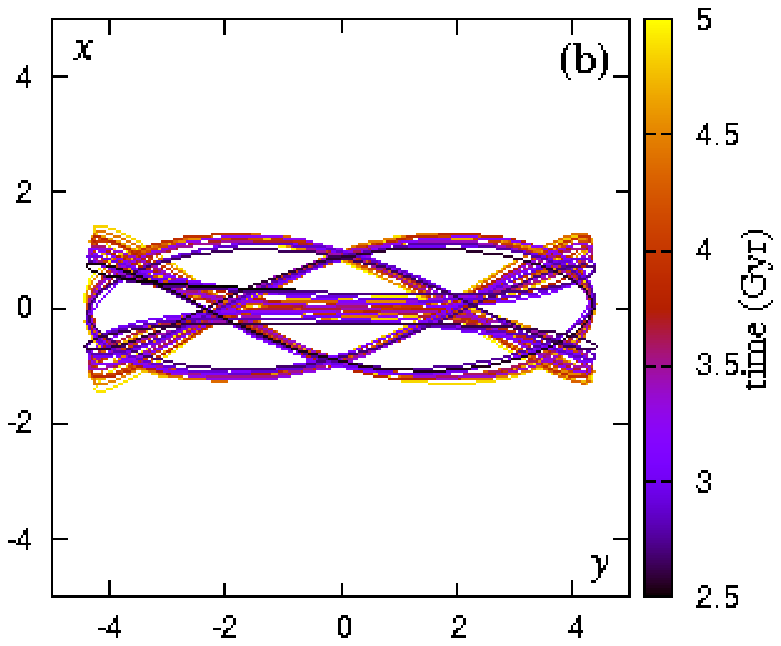}
\includegraphics[scale=0.475]{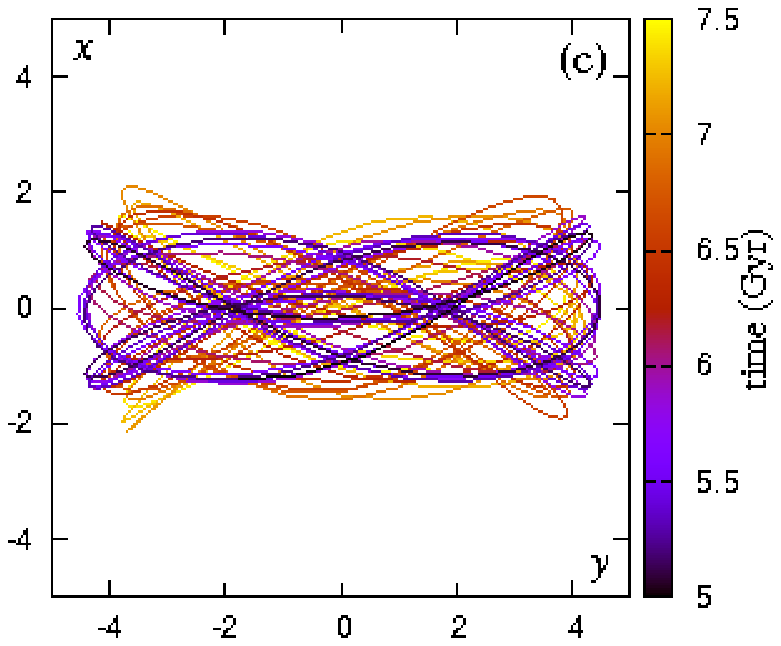}
\includegraphics[scale=0.475]{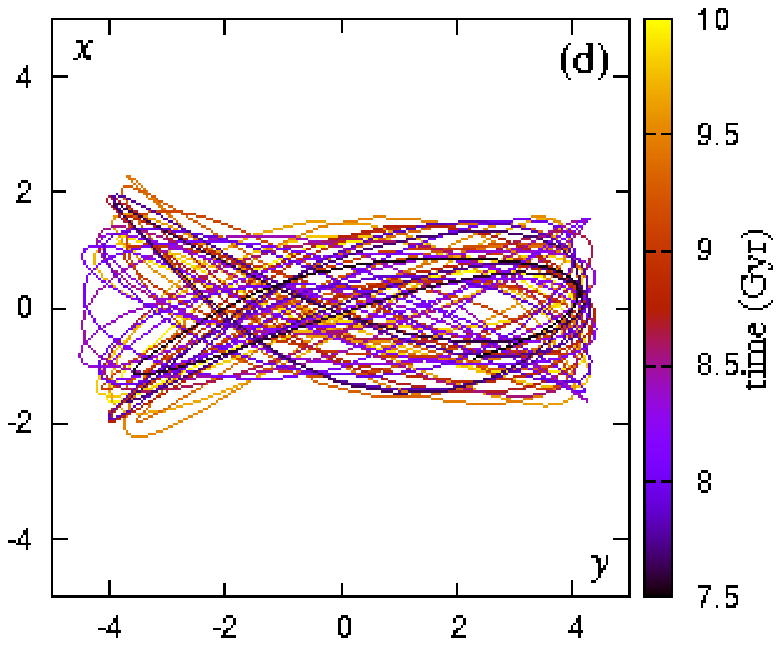}\vspace{-0.25cm}
\includegraphics[width=15.25cm,height=1.75cm]{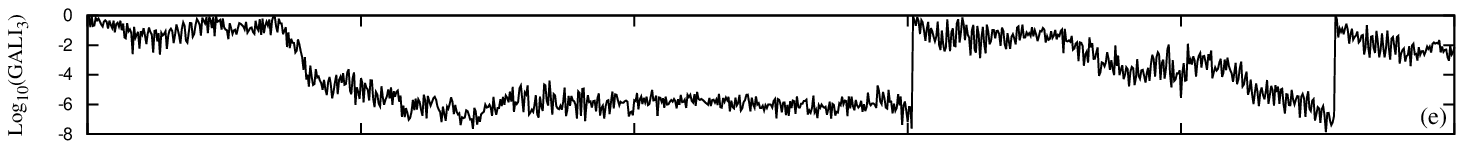}\vspace{-0.25cm}
\includegraphics[width=15.25cm,height=1.75cm]{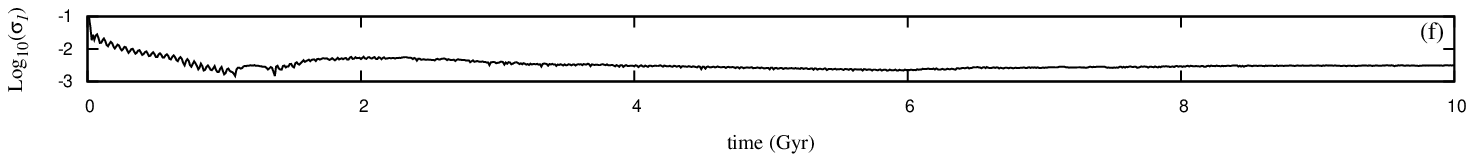}\vspace{-0.25cm}
\end{center}
\noindent\rule[0.25ex]{5cm}{0.5pt}
\begin{flushleft} \qquad \qquad \qquad  \textit{Scenario B} (bar-like orbit $B1$) \end{flushleft}
\begin{center}\hspace{0.9cm}
\includegraphics[scale=0.475]{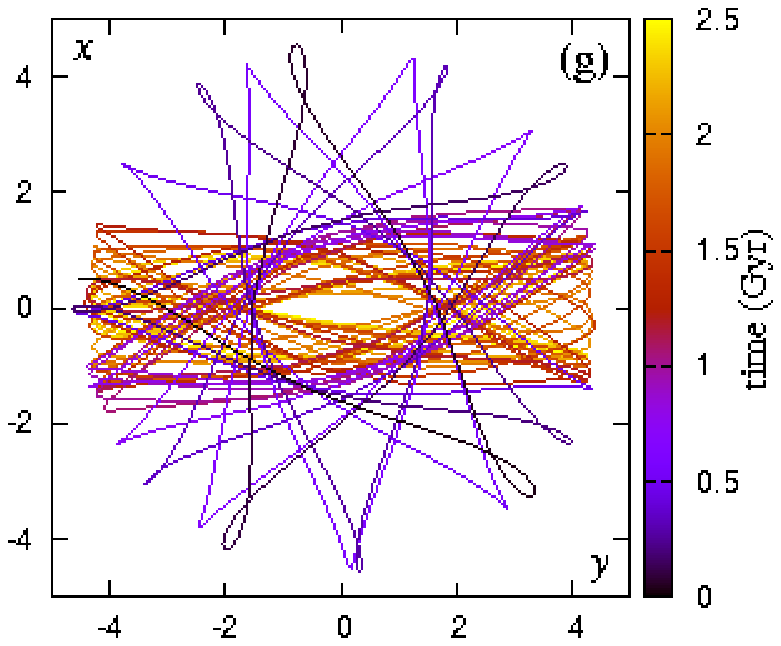}
\includegraphics[scale=0.475]{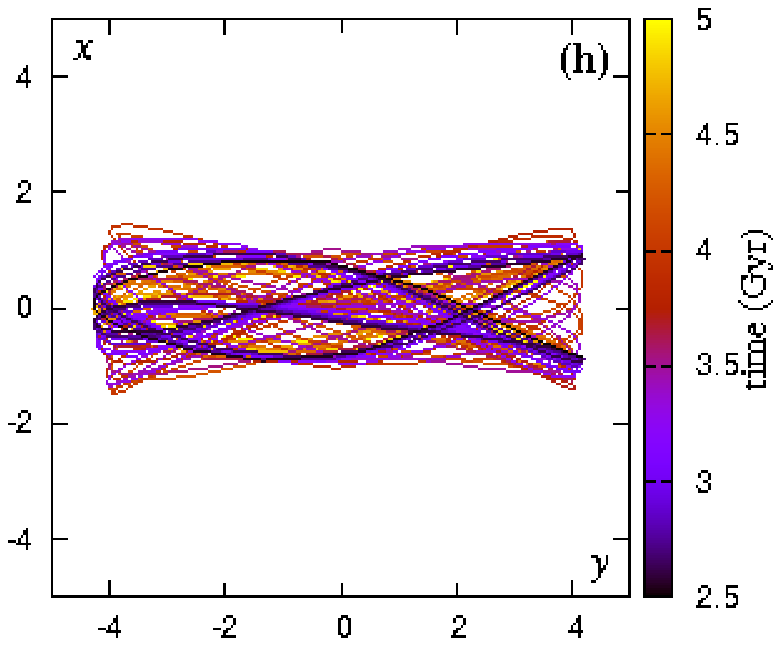}
\includegraphics[scale=0.475]{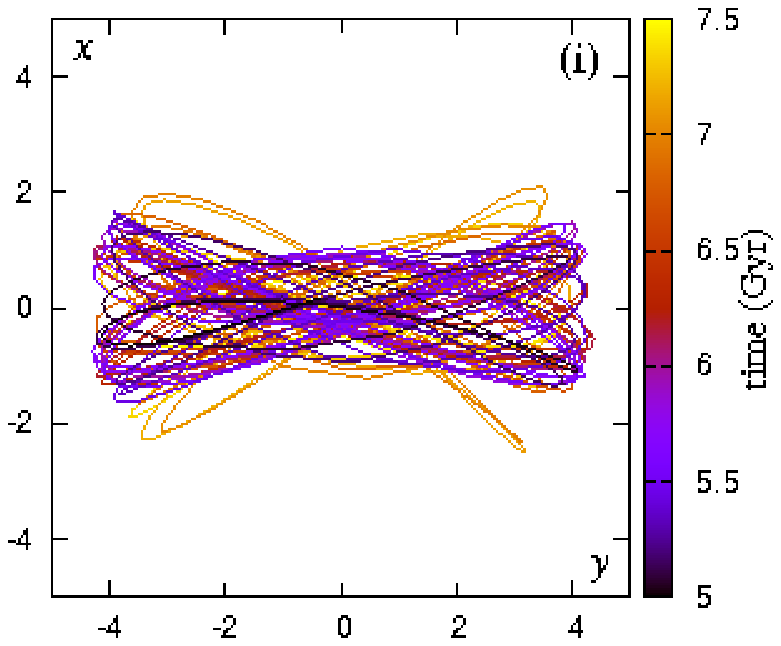}
\includegraphics[scale=0.475]{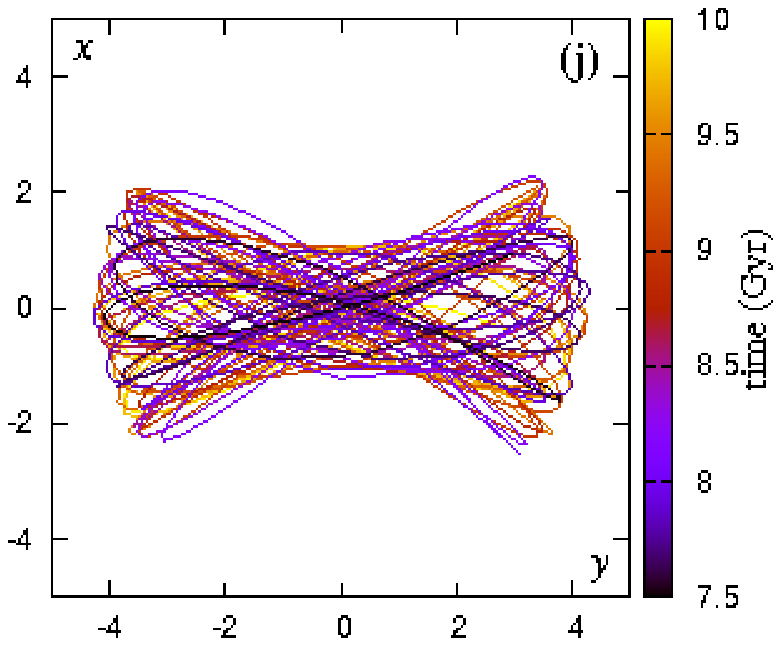}\vspace{-0.25cm}
\includegraphics[width=15.25cm,height=1.75cm]{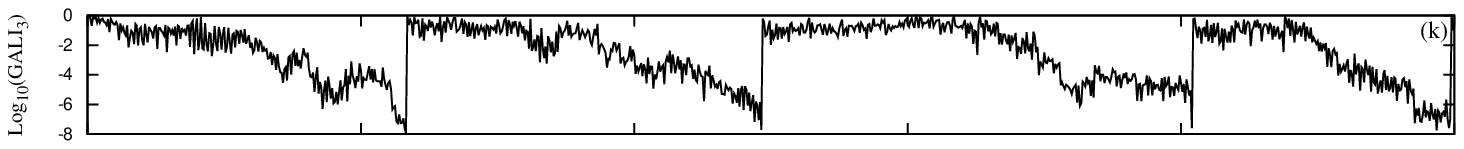}\vspace{-0.25cm}
\includegraphics[width=15.25cm,height=1.75cm]{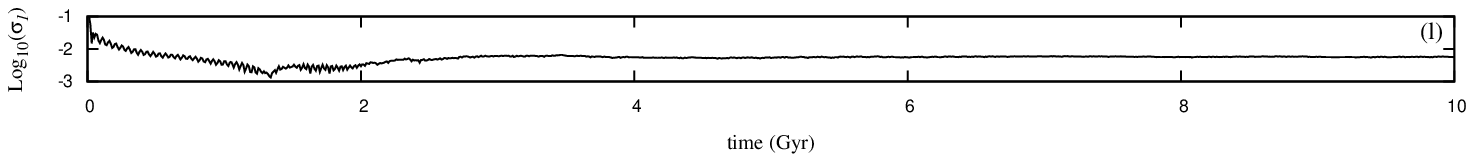}\vspace{-0.25cm}
\end{center}
\noindent\rule[0.25ex]{5cm}{0.5pt}
\begin{flushleft} \qquad \qquad \qquad  \textit{Scenario C} (bar-like orbit $B1$) \end{flushleft}
\begin{center}\hspace{0.9cm}
\includegraphics[scale=0.475]{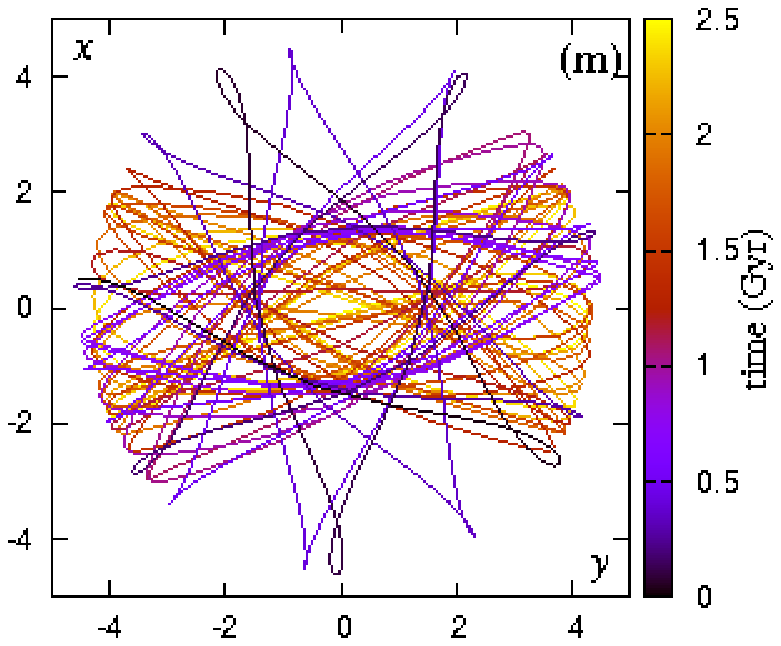}
\includegraphics[scale=0.475]{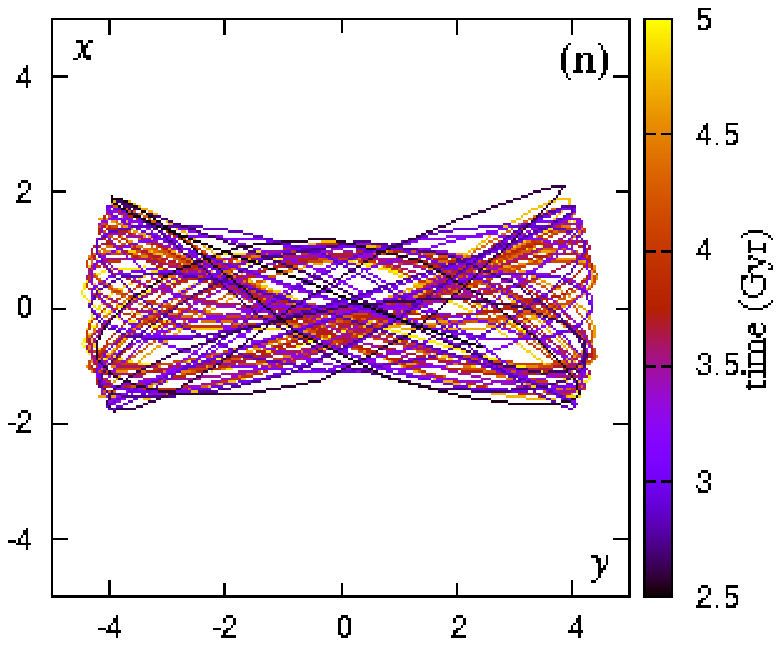}
\includegraphics[scale=0.475]{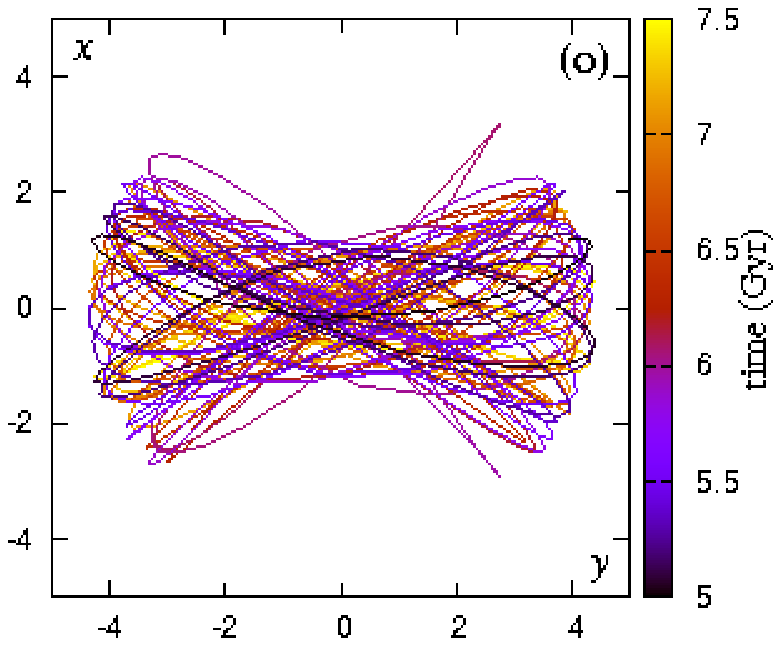}
\includegraphics[scale=0.475]{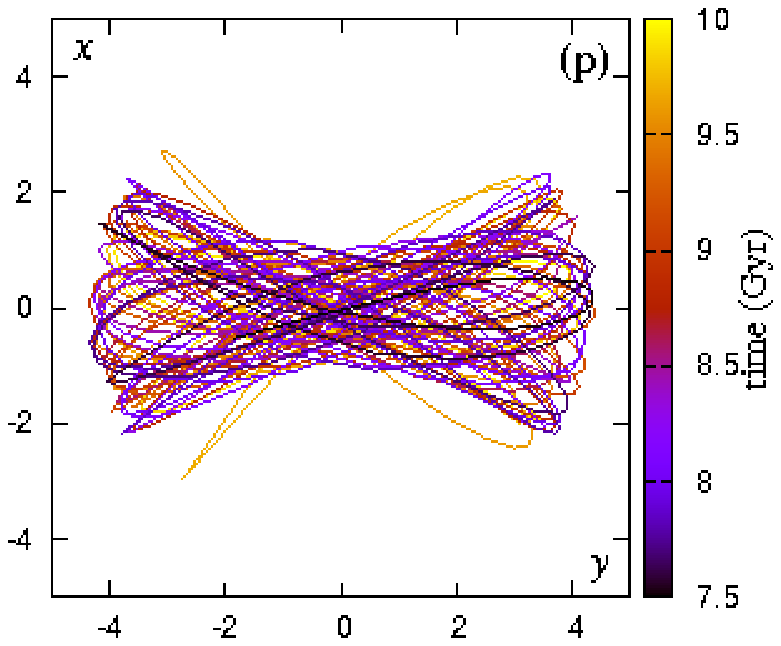}
\includegraphics[width=15.25cm,height=1.75cm]{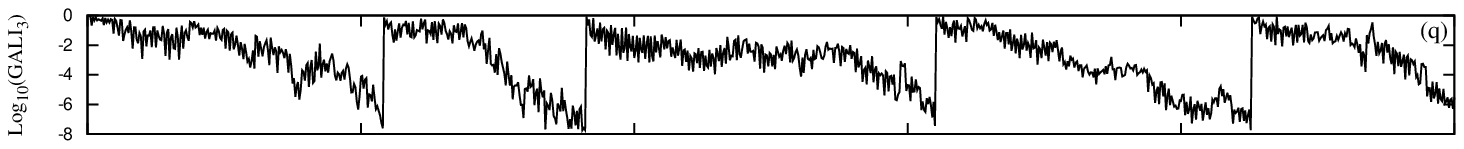}\vspace{-0.25cm}
\includegraphics[width=15.25cm,height=1.75cm]{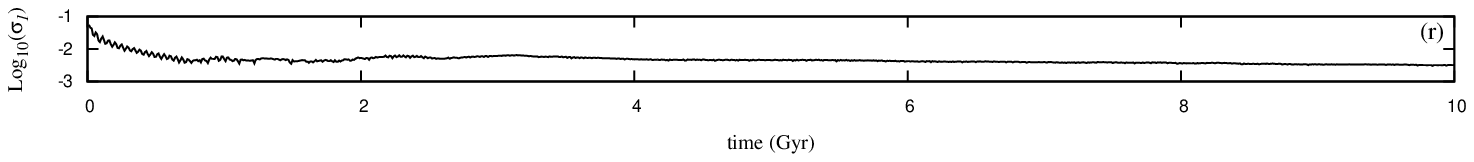}
\caption{(Colour online) The 3-d.o.f. orbit $B1$ evolved with the
\textit{Scenario A, B} and \textit{{C}}. Its different projection on the
$(x,y)$-plane in different time windows is depicted in the first (four-panel)
row (from top block-part to the bottom respectively) and the colour bar
corresponds to the time (in Gyr). Their GALI$_3$ and MLE $\sigma_1$ evolution
in time is shown for each case just below them. Note that the orbit (in all
three cases) starts as a disc-like and gradually its shape turns to barred,
displaying the bar's growth through the parameters of the Ferrers' potential
(see text for more details).} \label{fig:3Dbar}
\end{center}
\end{figure*}

\begin{figure*}
\begin{flushleft} \qquad \qquad \qquad  \textit{Scenario A} (disc-like orbit $D1$) \end{flushleft}
\begin{center}\hspace{0.9cm}
\includegraphics[scale=0.475]{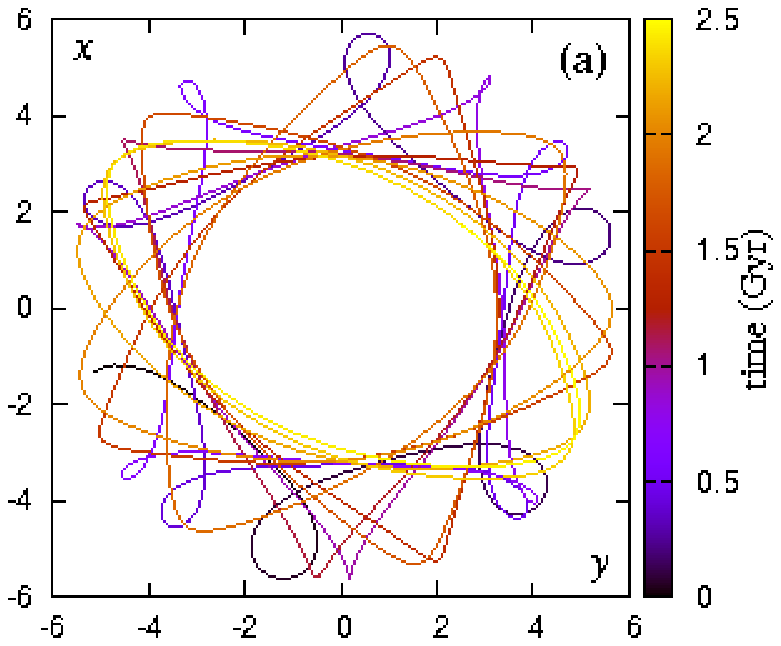}
\includegraphics[scale=0.475]{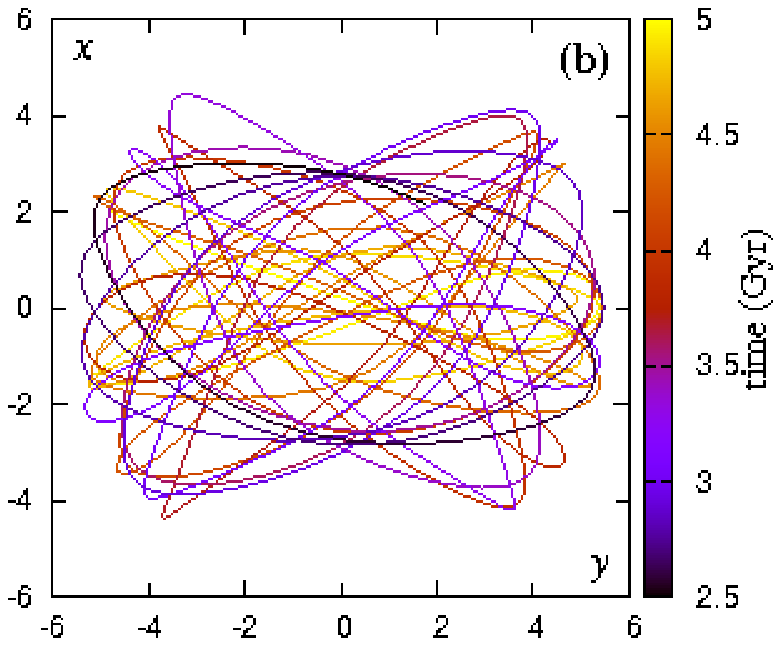}
\includegraphics[scale=0.475]{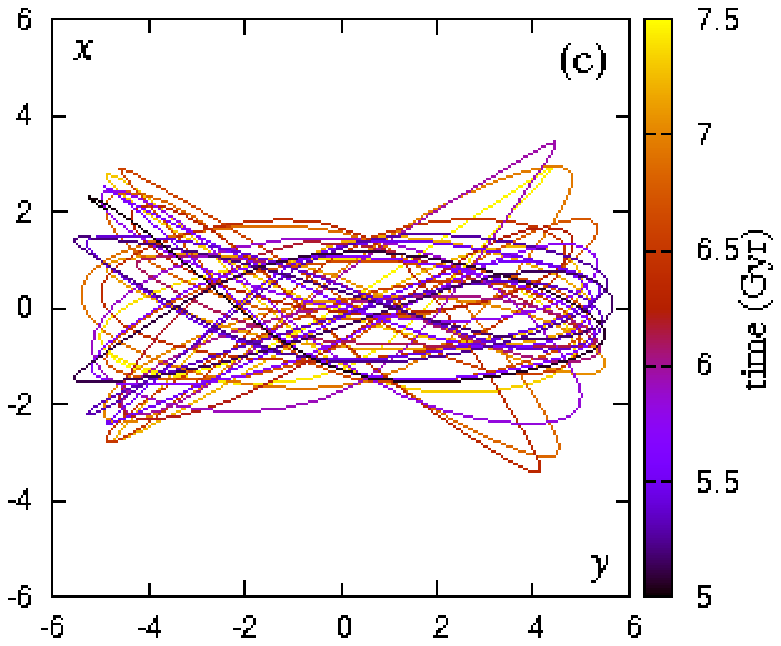}
\includegraphics[scale=0.475]{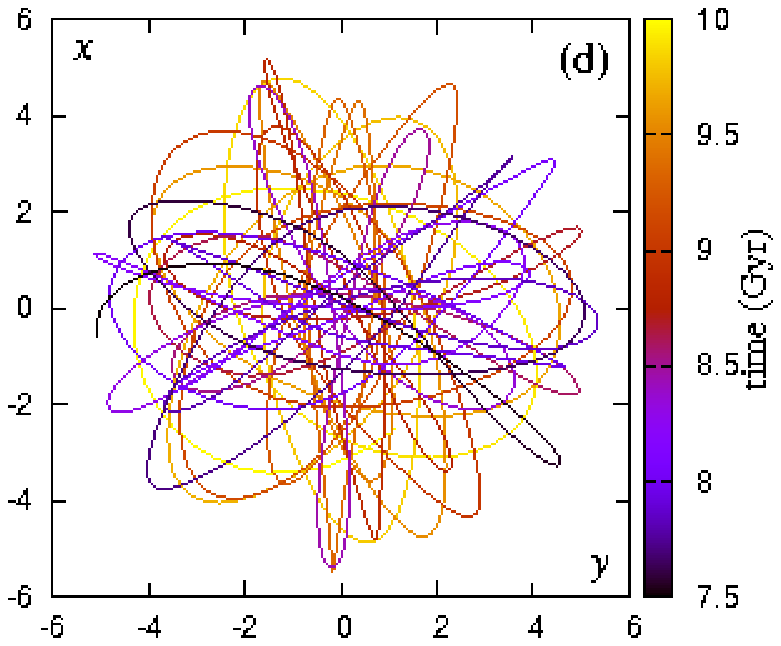}
\includegraphics[width=15.25cm,height=1.75cm]{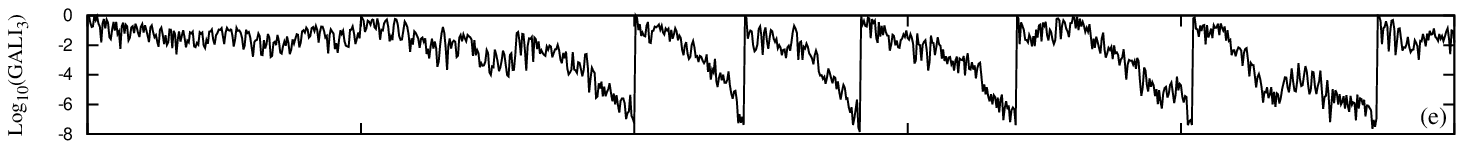}\vspace{-0.25cm}
\includegraphics[width=15.25cm,height=1.75cm]{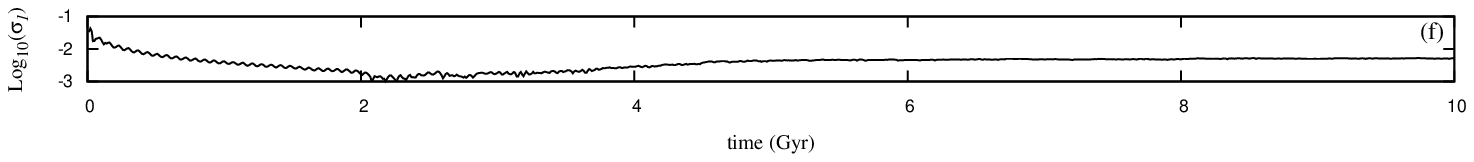}\vspace{-0.25cm}
\end{center}
\noindent\rule[0.25ex]{5cm}{0.5pt}
\begin{flushleft} \qquad \qquad \qquad  \textit{Scenario B} (disc-like orbit $D1$) \end{flushleft}
\begin{center}\hspace{0.9cm}
\includegraphics[scale=0.475]{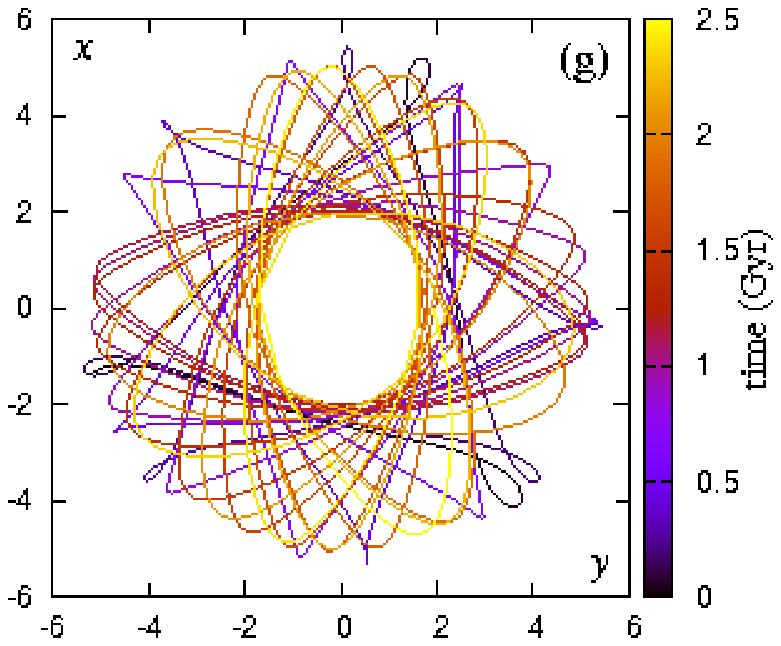}
\includegraphics[scale=0.475]{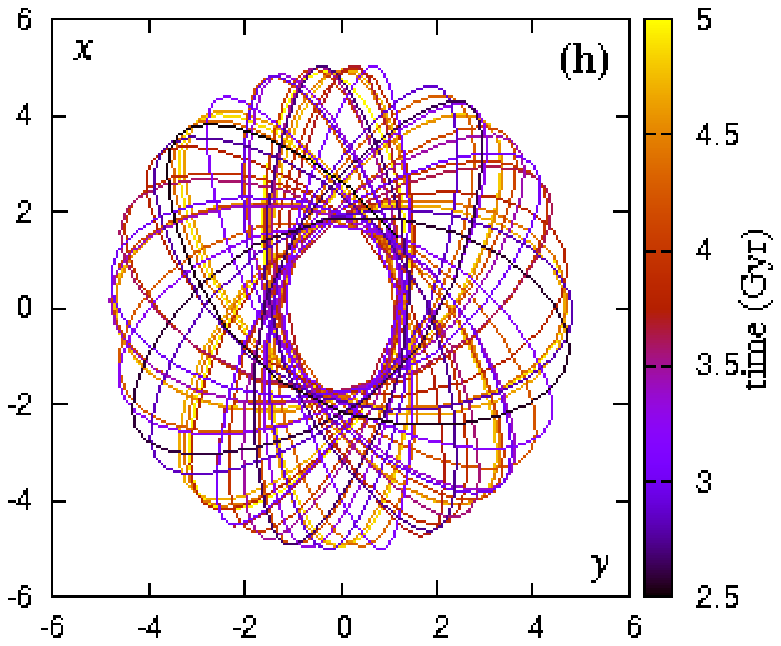}
\includegraphics[scale=0.475]{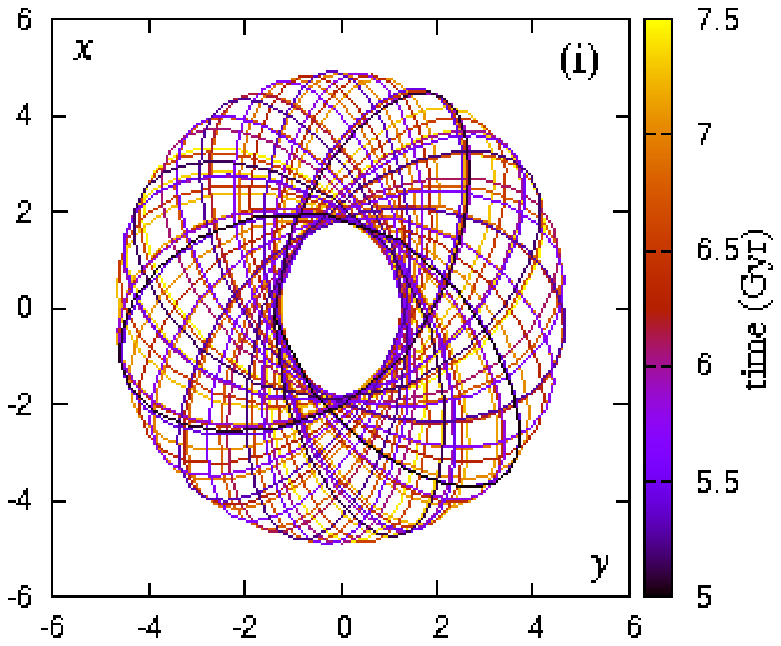}
\includegraphics[scale=0.475]{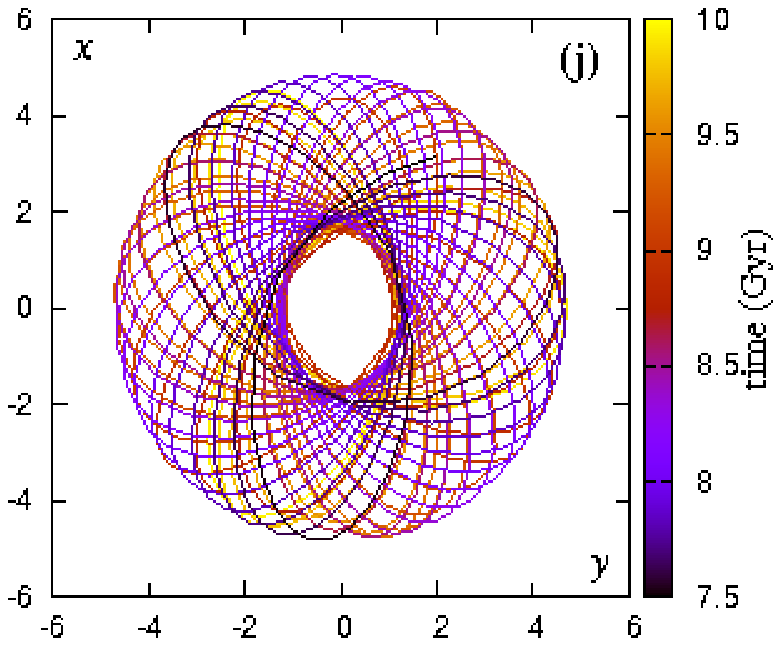}
\includegraphics[width=15.25cm,height=1.75cm]{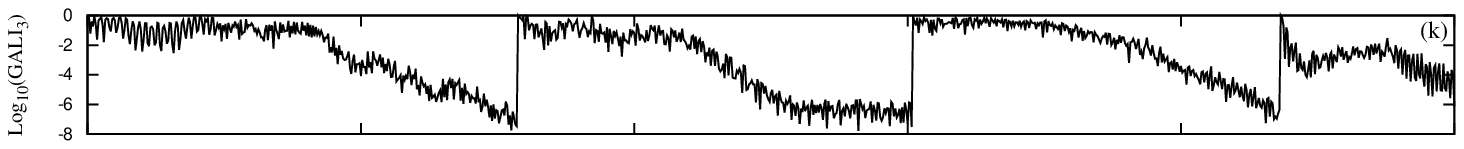}\vspace{-0.25cm}
\includegraphics[width=15.25cm,height=1.75cm]{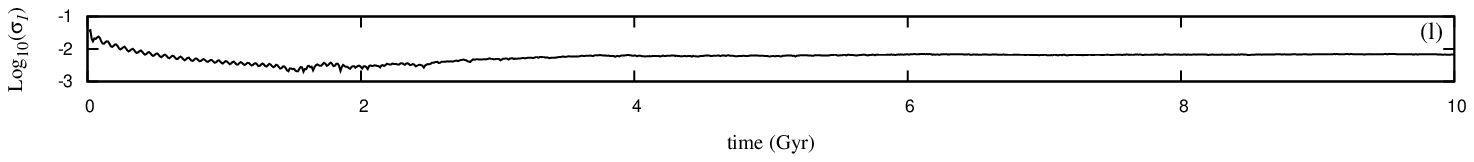}\vspace{-0.25cm}
\end{center}
\noindent\rule[0.25ex]{5cm}{0.5pt}
\begin{flushleft} \qquad \qquad \qquad  \textit{Scenario C} (disc-like orbit $D1$) \end{flushleft}
\begin{center}\hspace{0.9cm}
\includegraphics[scale=0.475]{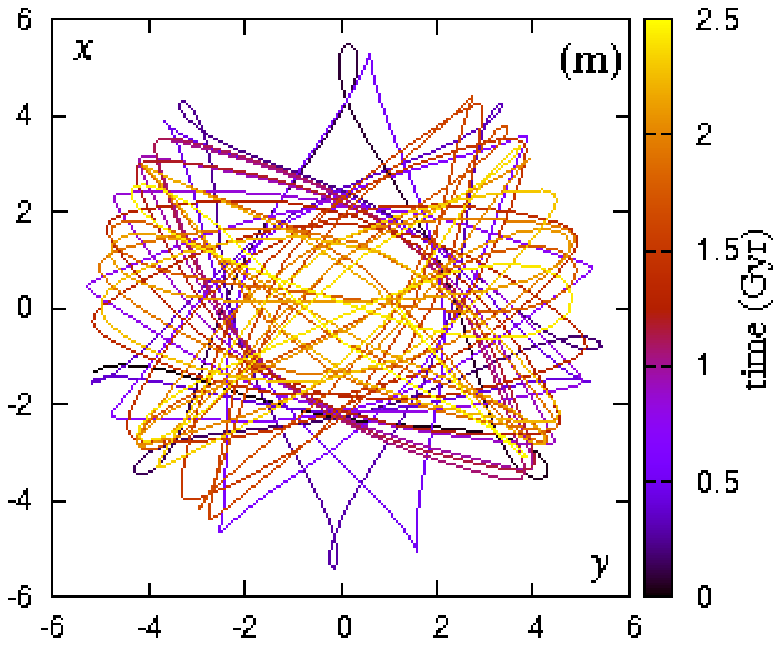}
\includegraphics[scale=0.475]{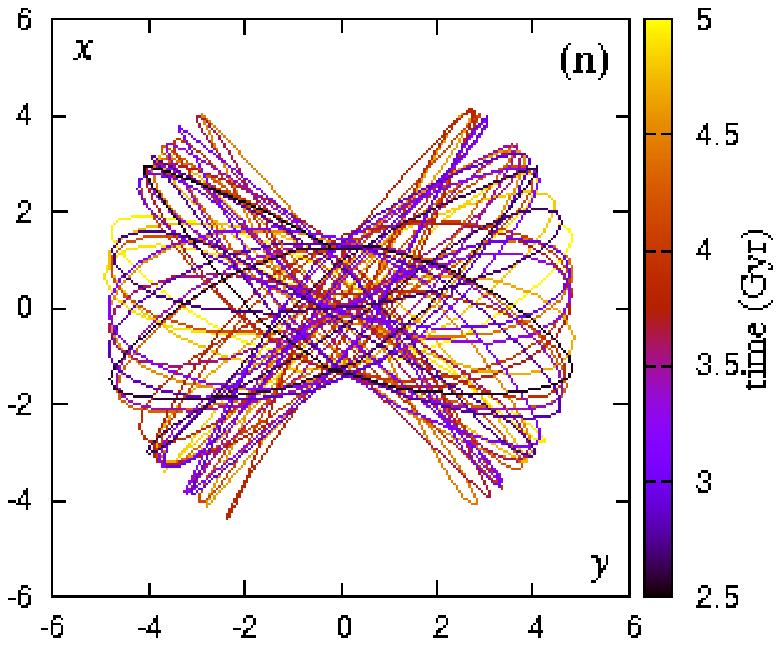}
\includegraphics[scale=0.475]{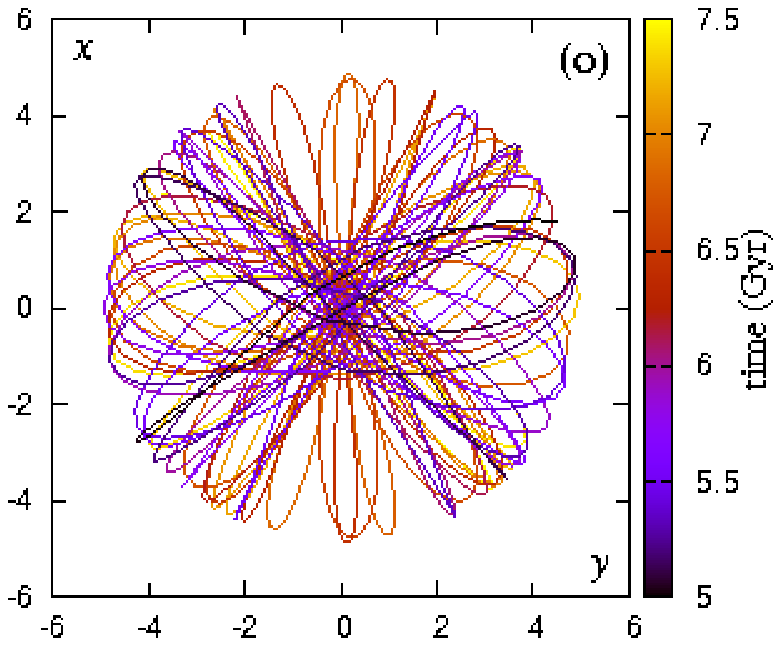}
\includegraphics[scale=0.475]{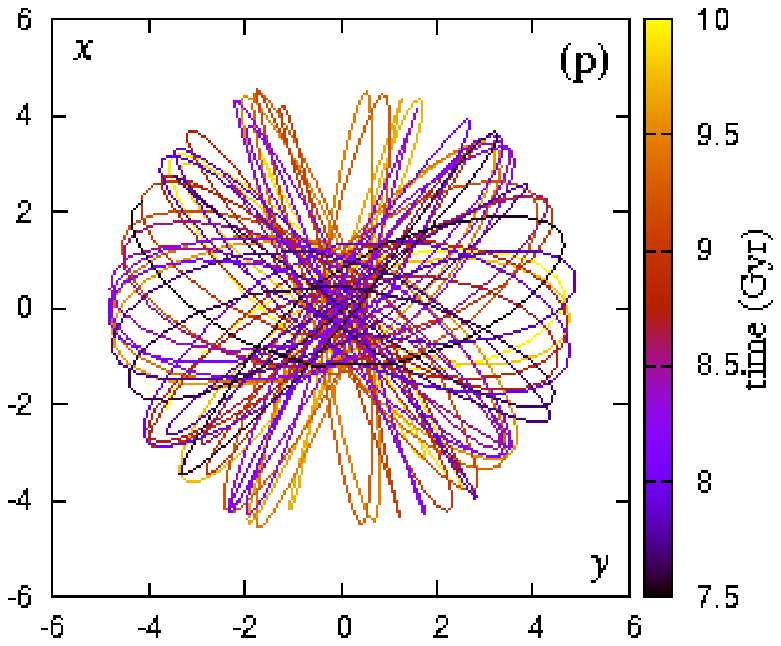}
\includegraphics[width=15.25cm,height=1.75cm]{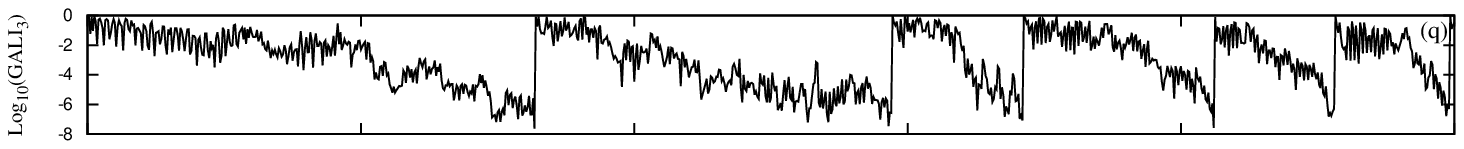}\vspace{-0.25cm}
\includegraphics[width=15.25cm,height=1.75cm]{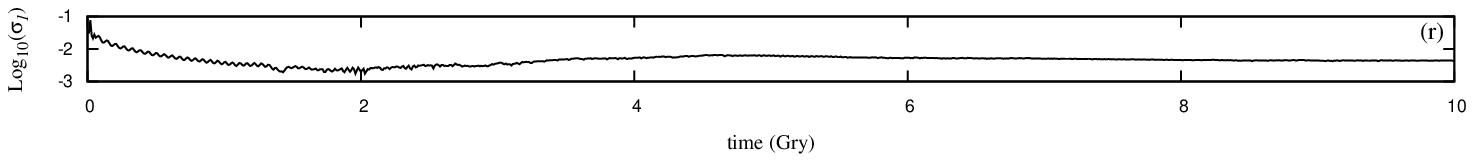}
\caption{(Colour online) Same as in Fig.~\ref{fig:3Dbar} but for the 3-d.o.f.
disc-like orbit $D2$ evolved again with the `Scenario A, B and C'. Note, that
here the disc-like pattern slightly varies from cases to case. The
\textit{different degree of chaoticity} can be accurately captured by the
frequency and fast decay to zero of the GALI$_3$
[Fig.~\ref{fig:3Ddisc}(e,k,q)], indicating that the orbits is relatively
`strongly chaotic' under the `Scenario A, C' while under the `Scenario B' is
relatively `weakly chaotic'. This information can not be revealed in such a way
by the MLE $\sigma_1$ shown in Figs.~\ref{fig:3Ddisc}(f,l,r) (see text for more
information).} \label{fig:3Ddisc}
\end{center}
\end{figure*}

Before focusing on the general dynamical trends of the above different
\textit{scenarios}, let us first study the evolution of a few typical examples
of orbits. We picked a random but representative (in terms of radial distance)
number of orbits from our 100~000 initial conditions and calculated their
orbital evolution together with their GALI$_3$ and MLE. Let us mention here
that for all three Scenarios the number of escaping orbits is negligible. From
this subset of orbits, we chose to show here, in Fig.~\ref{fig:3Dbar} and
Fig.~\ref{fig:3Ddisc}, two typical examples of trajectories which present also
a rather `rich' and interesting morphological behaviour. Each figure is divided
in three main blocks, each one corresponding to the different evolution
Scenarios A, B and C respectively. Then, each block is further composed of
three parts where the orbital evolution is depicted on the $(x,y)$-plane in the
course of time (first rows of each subpart), together with the GALI$_3$ and MLE
$\sigma_1$ in the next two rows respectively.


Thus, in Fig.~\ref{fig:3Dbar} we show the evolution of an orbit from the
ensemble of the $N$-body simulation with initial condition
$(x,y,z,p_x,p_y,p_z)\approx$~$(-4.543100,0.499639,-0.162627,\newline
0.048798,-0.218718,0.002898)$ (we will refer to this orbit from now on as
`$B1$'), which is iterated for 10~Gyr (10~000 time units) for the TD potentials
mentioned above. The $B1$ orbit belongs to a set of initial conditions which is
representative of the imposed scenario of the bar growth by our TD potential.
The starting and complete set of parameters for the TD model is taken at
$t_0=1.4$~Gyr by the fits with the $N$-body simulation with the procedure
described in Section~\ref{NbodyvsPot}. Note that in all figures' panels, we set
everywhere the $t_0$ equal to zero instead of $t_0=1.4$~Gyr.

In the first row of the first block in Fig.~\ref{fig:3Dbar}(a,b,c,d), we show
its projection on the $(x,y)$-plane for four successive time intervals of
$\Delta t = 2.5$~Gyr when evolved by the `Scenario A', i.e., all the three
potential components $V_B,V_D,V_H$ are time-dependent and the total energy is
not in general conserved. The colour bar next to each panel corresponds to the
time (in Gyr), hence the most recent epochs of the orbit are coloured with
yellow (light-grey in b/w) while those in the earlier ones with dark blue or
black (dark-grey or black in b/w). In Fig.~\ref{fig:3Dbar}(e) we show its
GALI$_3$, capturing accurately the chaotic nature of the orbit during the first
[Fig.~\ref{fig:3Dbar}(a)], third [Fig.~\ref{fig:3Dbar}(c)] and fourth
[Fig.~\ref{fig:3Dbar}(d)] time windows by decaying exponentially to zero. On
the other hand, in the second [Fig.~\ref{fig:3Dbar}(b)] time window its regular
(even by just looking its projected morphologically on the $(x,y)$-plane)
behaviour is successfully revealed by the fluctuates to a non-zero value of the
index. Note that the plot is in lin-log scale and the deviation vectors are
re-initialized, by taking again $k$ new random orthonormal deviation vectors,
each time the GALI$_3$ becomes very small (i.e.~GALI$_3 \leq 10^{-8}$). It
turns out that the $B1$ begins as a regular disc-like orbit during the first
2.5~Gyr and, as the bar starts forming and growing, it gradually evolves to a
chaotic bar-like orbit until the end of the integration. We may notice how hard
it is for the finite time MLE $\sigma_1$ [Fig.~\ref{fig:3Dbar}(f)] to capture
these different dynamical different transitions and epochs due to its
time-averaged definition \citep[see also][]{ManBouSkoJPhA2013}. Furthermore,
its power law decay for regular time intervals and its tendency to positive
values for chaotic ones are of the same order of magnitude making it rather
hard to use the temporary value $\sigma_1$ as a safe criterion of regular and
weak or strong chaotic motion.

In order to see how morphologically sensitive our `full' TD parameter model is
(as described in `Scenario A') to the several component parameters and the lack
of energy conservation, let us evolve the same initial condition ($B1$) with
the `Scenario B'. We can see that, the general shape of the orbits is more or
less similar, starting again as an disc-like orbit and drifting to a bar-like
one in time [first row of the second block in Fig.~\ref{fig:3Dbar}(g,h,i,j)].
Of course, the dynamical epochs are a bit different. In the first row of the
third block in Fig.~\ref{fig:3Dbar}(m,n,o,p), we show its evolution when
integrated with the time-dependencies described in the `Scenario C' and
recovering again similar morphological behaviour. In the elongated panels,
beneath the projections on the $(x,y)$-place of the `Scenario B, C', we depict
their corresponding GALI$_3$ [Fig.~\ref{fig:3Dbar}(k,q) respectively] and MLE
$\sigma_1$ [Fig.~\ref{fig:3Dbar}(l,r) respectively] just like for the `Scenario
A'.

Of course, we should not expect such a good agreement in the several
morphological behaviours in general and for all orbits of the initial ensemble.
More specifically, and in cases where chaos is strong from the very early
moments, the orbits are expected to differ significantly depending on the
evolutionary scenario. As also discussed in \cite{CarWac2006CeMDA}, the chaotic
behaviour is sensitive to the choice of the potential, even if it is frozen
like in their case.

In Fig.~\ref{fig:3Ddisc}, and in a similar manner as in Fig.~\ref{fig:3Dbar},
we show another characteristic disc-like orbit for most of the total of
integration with initial condition $(x,y,z,p_x,p_y,p_z)\approx$~$(-5.14416,
-1.345540, 0.277956,\newline 0.140120, -0.219648, 0.000338)$ (we will refer to
this orbit from now on as `$D1$'). The evolutionary scenarios are again the
same as before, i.e., in the first row of the first block in
Fig.~\ref{fig:3Ddisc}(a,b,c,d), we present its projection on the $(x,y)$-plane
for different time windows. The $D1$ orbit experiences a regular epoch during
its first 2.5~Gyr, then gradually becomes chaotic switching to a bar-like shape
and finally becomes a chaotic but disc-like now orbit. Its regular and chaotic
epochs are accurately captured by the GALI$_3$ [Fig.~\ref{fig:3Ddisc}(e)],
fluctuating to constant value for the first 2.5~Gyr and then successively
decaying exponentially to zero for the rest of the integration. The MLE
$\sigma_1$ [Fig.~\ref{fig:3Ddisc}(f)] also reveals this dynamical evolution, by
decaying with a power law for the regular part and converging to non-zero value
for the three last time windows. However, here the motion does not present any
further transition and/or interplay between regular and chaotic motion and
again (as for the $B1$ orbit) the order of magnitude for the $\sigma_1$ is not
varying sufficiently enough to lead to a safe conclusion at certain times
without seeing its whole time evolution.

Its evolution with the `Scenario B', is shown in the first row of second middle
block in Fig.~\ref{fig:3Ddisc}(g,h,i,j)] where it turns out that presents
different morphological shape. In more detail, it still begins similarly but
for later times it forms a disc-like orbit (less chaotic also) which does not
visit the central region compared to what happens in `Scenario A'. Moreover,
when looking at its shape when integrated with the `Scenario C'
[Fig.~\ref{fig:3Ddisc}(m,n,o,p)], we can again conclude that it is a chaotic
disc-like orbit but not identical to the other cases. Notice that, the
\textit{different degree of chaoticity} can be accurately captured by the
frequency and fast decay to zero of the GALI$_3$
[Fig.~\ref{fig:3Ddisc}(e,k,q)], indicating that the orbits is relatively
\textit{strongly chaotic} under the `Scenario A, C' while under the `Scenario
B' is relatively \textit{weakly chaotic}. This information can not be revealed
in such a way by the MLE $\sigma_1$ [Fig.~\ref{fig:3Ddisc}(f,l,r)].

Furthermore, let us clarify the following for such TD systems. When speaking
more strictly and using more rigorous notions from non-linear dynamics theory,
those orbits experiencing such an interplay between regular and chaotic epochs
in different time intervals are in general (asymptotically) chaotic. Their MLE
is positive (like in our examples in Fig.~\ref{fig:3Dbar} and
Fig.~\ref{fig:3Ddisc}). However, when one is interested in astronomical time
scales, one may split the time in smaller windows and see how an ensemble of
orbits evolve. There, one may detect the different orbital trends (general at
first), in terms of chaotic or regular by checking the \textit{local
exponential divergence}. The GALI method is one good way to capture these
phenomena quite successfully, after giving enough time to the deviation vectors
to detect the stability of the local dynamics. Since one cannot know or predict
in advance, when exactly such transitions between regular and chaotic motion
will occur for each individual trajectory, one is limited to the study of
rather average trends.

One common approach broadly performed in the recent literature is to measure
the relative fraction of regular and chaotic motion in a sequence of frozen
snapshots \citep[e.g.][]{ValDebQuiMoo2010MNRAS}. However, this would not allow
us to capture an important and abundant orbital behaviour in the $N$-body
simulations, namely orbits that alternate their nature from regular to chaotic
and vice-versa as well as their morphological shape, e.g. from disc-like shape
to barred-like one.

\begin{figure}
\begin{center}
\includegraphics[width=\columnwidth]{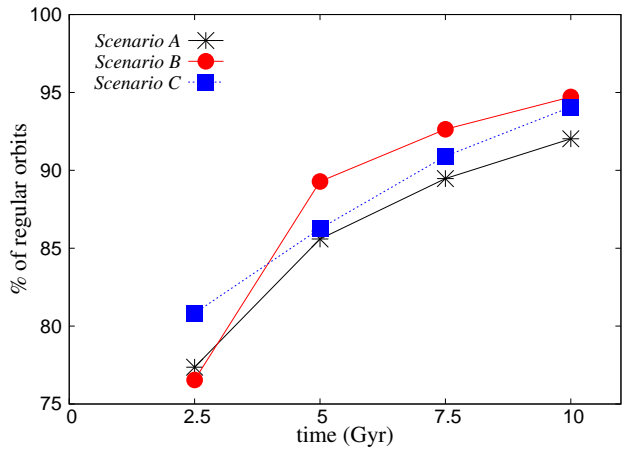}
\caption{(Colour online) Percentages of regular motion for different time
windows.} \label{3DOF_GALIpercICs}
\end{center}
\end{figure}

Turning now to the aspect of the global (in)stability of the model and this set
of initial conditions in particular, one would be interested to monitor the
variation of the total fraction of regular \textit{vs.} chaotic motion in the
course of time. Since our model is TD the percentage of chaotic orbits
following the potentials (described as Scenarios A, B and C) is expected to
change in time. In order to estimate this we adopt the following strategy: We
divide the total integration time of 10~Gyr in four successive time windows of
length $\Delta t=2.5$~Gyr time units. At the beginning of each time window, we
re-initialize GALI$_3$ to unity and follow a new set of three orthonormal
deviation vectors for each orbit. Then for each time window we calculate the
current percentage of regular (non-chaotic) orbits as the fraction of orbits
whose GALI$_3$ remains $> 10^{-8}$ during that time interval, i.e., GALI$_3$
does not decay exponentially (fast) to small values. In this way we allow the
GALI$_3$ to have enough evolutionary time to reveal the chaotic or regular
nature of the orbit within this interval. The results of this procedure are
shown in Fig.~\ref{3DOF_GALIpercICs},
where we depict the percentages of regular motion for different time windows
for the ensembles of initial conditions evolved under the three different
evolutionary processes. Though, it turns out that all three dynamical
evolutionary cases show quite similar trends, in terms of fraction of chaotic
and regular motion, implying that the variations of the halo and disc
parameters effect is relatively weak with respect to the bar's. The amount of
regular motion for the three different cases systematically grows in time,
starting from $\approx 75\%$ to $\approx 81\%$ after $2.5$~Gyr reaching
$\approx 90\%$ to $\approx 94\%$ after $10$~Gyr. The `Scenario A' turns out to
be slightly more chaotic compared to other two. Moreover, let us stress the
fact that the presence of regular motion is abundant in this TD model when
comparing to similar studies \citep{ManAthMNRAS2011,ManBouSkoJPhA2013} where
chaos turned out to be dominant.

The next step is to see how the bar strength is related to the general
(average) stability. As found and discussed in \cite{ManAthMNRAS2011}, and for
TI modes, one should expect to find a strong correlation between the strong
bars and the total amount of chaos. This result was confirmed for a simple TD
model, where only the mass of bar was considered to grow in time in
\cite{ManBouSkoJPhA2013}, in a potential also composed of a bar and a disc but
of a bulge instead a halo component. By simply looking at the percentages of
regular motion in Fig.~\ref{3DOF_GALIpercICs}, one would conjecture that the
relative non-axisymmetric forcings are decreasing slightly in time, implying
that the bar gets weaker.

However, this is not what happens here. As shown in Fig.~\ref{fig:A2}, the
$A_2$ parameter (maximum relative contribution of the $m=2$ Fourier component
of the mass distribution in the disc) increases in time during the first part
of the evolution  and stays more or less constant till the end, for both the
$N$-body simulation and the TD analytical model. In order to interpret
appropriately the increase  of regularity with the simultaneous increase of the
non-axisymmetric forcings, we should look back to the 2-d.o.f. case and the
PSSs in Fig.~\ref{fig:2Dgalipss}. There, one may see that for the relative
energy interval values of the ensemble of 3-d.o.f. initial conditions (chosen
from the $N$-body simulation) the \textit{stable island}, associated to the
bar-like orbits, gets larger in the course of time. Combining this with the
fact that both the self and non-self consistent models enhance the barred
morphological feature as time grows, we may conclude that the fraction of
regular motion increase in time is linked to the underlying growth of the
central `barred' island of stability. Moreover, it implies that these initial
conditions populate with higher probability this part of the (6 dimensional)
phase space of our 3-d.o.f. model which tends in time to encompass larger
regular area with bar-like orbits within.

Thus, it becomes evident that for a TD model the relative fraction of regular
(or chaotic) motion is not straightforwardly correlated to the bar's strength.
In order to understand such dynamical trends, one has to shed some light to the
underlying time-dependent dynamics. This can be done by considering how the
ensemble is distributed in the energy interval, and the specific dynamical
trends of the model. These trends are manifested by specific morphological
properties, such as growth of the `barred' stable island in our case. They may
also affect the global (in)stability in different manners depending on the
model.

\section{Summary and conclusions} \label{concl}

In order to carry out all the analyses summarized below, we employed an
analytical model that was specially tailored for this purpose. It was based on
the results of a self-consistent $N$-body simulation of an isolated barred
galaxy. We measured several structural parameters of this simulation, as a
function of time, and then used them to set up the analytical gravitational
potential of a galactic model. This model was composed of three components,
representing the disc, the bar and the dark matter halo. It implements
analytical potentials that are meant to be conveniently simple, while at the
same time being able to mimic the simulation to a reasonable degree of
accuracy. It should de pointed out that our analytical model is not based on
frozen potentials, in any way. Instead, it is a fully time-dependent model,
that relies upon the detailed features of the $N$-body simulation. This entails
that, contrary to the frozen (and more accurate) ones, where the variety of
orbital motion is restricted and destined to be either regular or chaotic, our
TD model is equipped with some extra orbital behavior, namely trajectories
which may alternate nature, behaving regularly for some epochs and chaotically
for others, and in a not necessarily in a monotonic way. This is also what
broadly happens in the $N$-body simulations.

The adequacy of the model we constructed is verified in at least three
essential ways. First, the similarity of the rotation curves ensures the global
dynamics should be well approximated. Second, if we compute the orbits of an
ensemble of test particles subject to this potential, they give rise to
morphological disc and bar features remarkably similar to those of the $N$-body
simulation. Third, the length and strength of the bar in the resulting mock
snapshots (from the analytical model) are in very good quantitative agreement
with the $N$-body bar. Such comparisons indicate that our model is able to
adequately capture both the dynamics and the morphology of the barred galaxy
model in question.

Starting with the reduced 2-d.o.f. and \textit{time-independent} case of the
model, we used the GALI method to successfully survey the underlying dynamics
of the phase space by mapping the chaotic and regular regimes (and motion). By
measuring these, we found that the fraction of regular orbits increases in
time, i.e., for model parameters at later times, for almost all energies.
Moreover, the island of stability associated to the bar-like trajectories also
gets larger in time, implying that the bar features are enhanced as time
evolves. This tasks allowed us to get a brief idea also for the full 3-d.o.f.
\textit{time-dependent} as well as the $N$-body simulation, exhibit a bar
growth evolution.

Regarding this TD analytical model, we similarly estimated \textit{stability
trends} in terms of estimating the amount of regular and chaotic motion in
different time-windows. In this case, we used a more realistic set of initial
conditions coming directly from the simulation itself and iterated them under
the constructed TD potential. In order to do that we firstly monitored the
GALI's detection efficiency to a representative sample of orbits and presenting
in this paper two of them which show some typical generic behaviour. We also
discussed its advantages with the traditional MLE in capturing such sudden
dynamical variations. Of course we did not manage to span the whole orbital
richness of our ensemble of initial conditions but we have rather achieved to
give a flavor of the possible evolutions for individual trajectories. It turns
out that the complete set of orbits tends to become relatively more `regular'
in time. To further examine this, and since now the total energy is not
conserved, we tried two alternative but similar models whose Hamiltonian
function value can be conserved by adjusting the disc's and/or the halo
parameters appropriately. In all cases the trend was found to be the same.

Moreover for this 3-d.o.f. TD model presented here (`Scenario A'), we envisage
that the detailed study of orbital dynamics can be extended further by, for
example, classifying the disc and bar orbits from a morphological point of view
as the time varies. Additionally we can do the same for the other two
`Scenarios' described in Section~\ref{TDmodel}. Furthermore, we have already a
quite large ensemble of orbits to study diffusion properties in the phase (also
in configuration) space, just like in other papers in the literature, and
compare it with the one observed in the $N$-body simulation.

Approaches such as this are potentially suited to a broad class of applications
in galactic dynamics (e.g. double bars, central mass concentrations and even
bar dissolution). Once the output of an $N$-body simulation has been modeled
into a time-dependent analytical potential, a variety of analyses could then be
undertaken, particularly regarding orbital studies. Work that generally relied
on highly simplified (and usually frozen) analytical potentials could take
advantage of more astrophysically realistic galaxy models. This
\textit{bridging} would afford an approximation to the richness of detail of an
$N$-body simulation, at a lower computational cost and with the versatility of
a simple analytical formulation.

\section*{Acknowledgements}

We would like to thank Ch. Skokos, T. Bountis and M. Romero-G{\'o}mez, for
their comments and fruitful discussions. REGM acknowledges support from FAPESP
(2010/12277-9). TM was supported by the Slovenian Research Agency (ARRS) and
partially by a grant from the Greek national funds through the Operational
Program `Education and Lifelong Learning' of the National Strategic Reference
Framework (NSRF) - Research Funding Program: THALES, Investing in knowledge
society through the European Social Fund. This work has made use of the
computing facilities of the Laboratory of Astroinformatics (IAG/USP,
NAT/Unicsul), whose purchase was made possible by the Brazilian agency FAPESP
(grant 2009/54006-4) and the INCT-A. We also acknowledge support from FAPESP
via a Visiting Researcher Program (2013/11219-3), and we would like to thank
the Institut Henri Poincar\'e for its support through the `Research in Paris'
program and hospitality in the period during which part of this work took
place.

\bibliographystyle{mn2e}
\bibliography{ManMachMNRAS}
\bsp
\end{document}